\documentclass[aps,prd,10pt,twocolumn,superscriptaddress,nofootinbib
  ,noshowpacs,preprintnumbers]{revtex4-1}

\pdfoutput=1

\usepackage{amsmath}
\usepackage[utf8]{inputenc}
\usepackage{multirow}
\usepackage{graphicx}
\usepackage{latexsym}
\usepackage[colorlinks]{hyperref}
\hypersetup{allcolors=[rgb]{0,0,1}}
\usepackage[dvipsnames]{xcolor} 
\usepackage{xfrac}

\graphicspath{{Final_Fig/}}

\begin{document}

\title{Neutrino emission characteristics of black hole formation\\ in three-dimensional simulations of stellar collapse}

\author{Laurie Walk}
\affiliation{Niels Bohr International Academy and DARK, Niels Bohr Institute, University of Copenhagen, Blegdamsvej 17, 2100, Copenhagen, Denmark}
\author{Irene Tamborra}
\affiliation{Niels Bohr International Academy and DARK, Niels Bohr Institute, University of Copenhagen, Blegdamsvej 17, 2100, Copenhagen, Denmark}
\author{Hans-Thomas Janka}
\affiliation{Max-Planck-Institut f\"ur Astrophysik,
 Karl-Schwarzschild-Str.~1, 85748 Garching, Germany}
 \author {Alexander Summa}
\affiliation{Max-Planck-Institut f\"ur Astrophysik,
 Karl-Schwarzschild-Str.~1, 85748 Garching, Germany}
  \author {Daniel Kresse}
\affiliation{Max-Planck-Institut f\"ur Astrophysik,
 Karl-Schwarzschild-Str.~1, 85748 Garching, Germany}
 \affiliation{Physik-Department, Technische Universit\"at M\"unchen, James-Franck-Stra{\ss}e 1, 85748 Garching, Germany}
 
\date{\today}
\begin{abstract}

Neutrinos are unique probes of core-collapse supernova dynamics, especially in the case of black hole (BH)  forming stellar collapses, where the electromagnetic emission may be faint or absent. By investigating two 3D hydrodynamical simulations of BH-forming stellar collapses of mass $40$ and $75\ M_\odot$, we identify the physical processes preceding BH formation through neutrinos, and forecast the neutrino signal expected in the existing IceCube and Super-Kamiokande detectors, as well as  in the future generation DUNE facility. Prior to the abrupt termination of the neutrino signal corresponding to BH formation, both models develop episodes of  strong and long-lasting activity by the spiral standing accretion shock instability (SASI). We find that the spiral SASI peak in the Fourier power spectrum of the neutrino event rate will be distinguishable at $3\sigma$ above the detector noise for distances up to $\sim \mathcal{O}(30)$~kpc in the most optimistic scenario, with IceCube having the highest sensitivity. Interestingly, given the long duration of the spiral SASI episodes, the spectrograms of the expected neutrino event rate carry clear signs of the evolution of the blue spiral SASI frequency as a function of time, as the shock radius and post-shock fluid velocity evolve.  Due to the high accretion luminosity and its large-amplitude SASI-induced modulations,  any contribution from  asymmetric (dipolar or quadrupolar) neutrino emission associated with the lepton emission self-sustained asymmetry (LESA)  is far subdominant in the neutrino signal.

\end{abstract}

\maketitle

\section{Introduction} 

Core-collapse supernovae (SNe) occur when stars with a zero-age main sequence mass roughly between $10$--$150\ M_\odot$ end their lives with the onset of  gravitational collapse of their inner core~\cite{Fryer:1999mi, Colgate:1966ax,Colgate_1971,Woosley:2002zz}. The electron-degenerate iron core bounces and induces a shock wave in the infalling stellar mantle. As it travels outwards, the shock wave loses energy by dissociating iron nuclei, and stalls. According to the delayed SN mechanism, the SN then enters the accretion phase, in which infalling matter continually accretes onto the shock front~\cite{Bethe:1990mw}. In successful SNe, the shock wave is thought to be revived by neutrinos~\cite{Bethe:1984ux}. During the accretion phase, hydrodynamical instabilities such as the standing accretion shock instability (SASI), neutrino-driven convection, and the lepton emission self-sustained asymmetry (LESA) can develop, leading to large-scale asymmetries of mass distribution visible in the neutrino emission~\cite{Blondin:2002sm, Blondin:2006yw,Scheck:2007gw,Iwakami:2008qj, Marek:2007gr,Foglizzo:2011aa, Fernandez:2010db,Tamborra:2014aua}. Convective overturn and SASI also enhance the rate of neutrino heating, aiding the revival of the SN explosion. 

A revived SN explosion, in which  the stellar mantle is successfully ejected,  will result in the formation of a neutron star. However,  a black hole (BH) will be the outcome if the explosion mechanism fails, and matter continues to accrete onto the transiently stable proto-neutron star (PNS), pushing it over its mass limit. In $10$--$30\%$ of  all core-collapse cases,  or possibly more,   the massive star is  thus expected to end in a BH, as signified by the observation of disappearing red giants~\cite{Adams:2016ffj,Adams:2016hit,Smartt:2015sfa,Gerke:2014ooa} and foreseen by recent theoretical work~\cite{Ugliano:2012kq,Sukhbold:2015wba,OConnor:2010moj}.  Since the BH is expected to form only after a period of post-bounce accretion that continues for  a few fractions of a second up to several seconds, the neutrino signal is expected to terminate abruptly~\cite{1988ApJ...334..891B}. Black hole forming stellar collapses (sometimes also named ``failed SNe'') emit a faint electromagnetic signal originating from the stripping of the hydrogen envelope~\cite{Kochanek:2013yca,Lovegrove:2013hox,1980Ap&SS..69..115N}. Sometimes, the BH formation may occur with a considerably longer delay, if not enough energy is released to finally unbind the star. In this case, a fraction of the stellar matter will fall back onto the PNS within minutes to hours and the PNS will be pushed beyond its limit, leading to BH formation~\cite{Wong:2014tca,Colgate_1971}. These so-called ``fallback SNe'' represent an intermediate class between ordinary SNe and  BH-forming stellar collapses. The signals from fallback SNe  will be similar to those of ordinary SNe with possible additional sub-luminous electromagnetic displays~\cite{Zhang:2007nw}. Given   the possibly dim or absent  emission, BH-forming stellar collapses may be difficult to detect electromagnetically. Neutrinos (and possibly gravitational waves) can therefore be unique probes of these cataclysmic events. In this work, we will focus on the neutrino emission properties from BH-forming stellar collapses, where the BH formation occurs after  few hundreds of milliseconds.

Early 1D hydrodynamical simulations of BH-forming models in spherical symmetry were carried out 
in~\cite{Liebendoerfer:2002xn,Sumiyoshi:2007pp, Sumiyoshi:2008zw, Ugliano:2012kq,Sukhbold:2015wba,Fischer:2008rh}. In~\cite{OConnor:2010moj} the effects of the nuclear equation of state (EoS), mass, metallicity, and rotation on BH formation were studied with a large set of so-called 1.5D progenitor models (i.e., employing spherical symmetry and an ``approximate angle-averaged" rotation scheme). Recently, \cite{Pan:2017tpk} investigated the time dependence of BH formation on the nuclear EoS   in a 2D self-consistent simulation of a non-rotating $40\ M_\odot$  model. Together, these studies provide fundamental predictions of the key features of the neutrino emission properties from BH-forming models.  

The detection of a neutrino burst from a nearby BH-forming stellar collapse  is bound to yield precious hints on the BH formation, see e.g.~\cite{Yang:2011xd,Mirizzi:2015eza}. Moreover, the neutrino signal from BH-forming stellar collapses is of relevance in the context of the detection of the diffuse SN neutrino background; the detection of the latter  will indeed  provide insight on the fraction of BH-forming stellar collapses  on cosmological scales~\cite{Lunardini:2009ya,Keehn:2010pn,Nakazato:2015rya,Nakazato:2013maa,Priya:2017bmm,Horiuchi:2017qja,Moller:2018kpn}. 

The information carried by neutrinos from BH-forming stellar collapses has so far been explored using  inputs from 1D hydrodynamical simulations only. In recent years, state-of-the-art 3D simulations of core-collapse SNe with sophisticated energy-dependent neutrino-transport have highlighted the plethora of information carried by neutrinos, especially for what concerns the pre-explosion dynamics~\cite{Tamborra:2014hga, Tamborra:2013laa, Lund:2012vm,Kuroda:2017trn, Lund:2010kh,Walk:2018gaw, Walk:2019ier,Tamborra:2014aua,Vartanyan:2018iah,Burrows:2019rtd,OConnor:2018tuw,Mueller:2012ak,Ott:2012mr,Takiwaki:2016qgc,Muller:2017hht,Ott:2017kxl}. The first 3D simulation of  BH formation and fallback was carried out in~\cite{Chan:2017tdg,Muller:2019upo}, motivated by the interest to explain astrophysical observations of metal-poor stars, natal kicks, and spins of BHs. It was found that the mass accretion rate remains large even directly after a possible shock revival. The rapid growth of the PNS can also cause a contraction of the post-shock layer, which induces violent SASI activity~\cite{Blondin:2002sm, Mueller:2012ak,Janka:2016fox}, followed by the eventual collapse of the transient PNS into a BH~\cite{Chan:2017tdg,Kuroda:2018gqq,Pan:2017tpk}. 

This work aims to identify the dominant features in the observable neutrino signal characterizing BH formation. Our analysis is carried out with two 3D simulations of BH-forming stellar collapse with different progenitor masses and metallicities. The resulting neutrino emission properties are compared and contrasted to a set of simulations of ordinary  non-rotating core-collapse SNe previously investigated by this group (specifically, we compare to a suite of $11.2, 15, 20, \mathrm{and}\; 27\ M_\odot$ SN models, see \cite{Tamborra:2013laa,Tamborra:2014aua,Tamborra:2014hga,Walk:2019ier,Walk:2018gaw}). By directly comparing to this set of models, the neutrino properties which are unique to the post-bounce evolution of a star collapsing  to a BH are determined. We provide a first attempt to infer the underlying physical processes governing BH formation, using  the features detectable through neutrinos with the IceCube Neutrino Telescope~\cite{Abbasi:2011ss}, Super-Kamiokande~\cite{Ikeda:2007sa}, and DUNE~\cite{Acciarri:2015uup}. 

This paper is outlined as follows. In Sec.~\ref{sec:Simulations} the key features of the BH-forming  models used throughout this work are outlined. Section~\ref{sec:Properties} discusses the neutrino emission properties;  how the latter relate to the SASI and LESA hydrodynamical instabilities is discussed in Secs.~\ref{sec:Evolution_SASI}  and \ref{sec:LESA}, respectively.  The features detectable  in the neutrino signal which are unique to BH formation are presented in  Sec.~\ref{sec:Detectable}. Finally, conclusions follow in Sec.~\ref{sec:Conclusions}.

\section{Simulations of Black Hole Forming Stellar Collapses}\label{sec:Simulations}

In this Section, we briefly describe the key features of the two 3D hydrodynamical  simulations used throughout this work. The neutrino emission properties will be described in the next Section. 

Two simulations of a non-rotating  solar-metallicity ($Z_\odot \sim 0.0134$) $40\ M_\odot$ progenitor~\cite{Woosley:2007as} and a metal-poor ($Z \sim 10^{-4} Z_\odot$) $75\ M_\odot$  star~\cite{Woosley:2002zz} were carried out with the \textsc{Prometheus-Vertex} code \cite{Rampp:2002bq}, which includes three neutrino flavors, energy dependent ray-by-ray-plus neutrino transport, and state-of-the-art modeling of the microphysics~\cite{Marek:2007gr,Buras:2005tb,Rampp:2002bq,Mueller:2012is}. Other codes employ   multi-D two-moment methods for the neutrino transport, see e.g.~\cite{Skinner:2018iti,Glas:2018oyz,Couch:2013kma}. 
In~\cite{Glas:2018oyz}, 3D core-collapse simulations of a non-exploding 20\,$M_\odot$ model and an exploding 9\,$M_\odot$ model were compared with ray-by-ray-plus transport approximation and multi-D transport. No major differences with respect to the hydrodynamic evolution and the basic features of the neutrino emission (luminosities and mean energies as functions of time) could be found beyond the small-scale (in space and time) fluctuations that are characteristic of the ray-by-ray method and variations in the non-linear dynamics that generically exhibit stochasticity and can be triggered by any kind of differences between simulations (e.g., a trivial phase shift of large-amplitude SASI shock oscillations in the 20\,$M_\odot$ model). Naturally, however, from these two representative models it cannot be concluded that cases do not exist where method-specific transport effects can be more relevant, for example in the collapse of rapidly rotating stellar cores, where the global deformation of the proto-neutron star sets limitations to the applicability of the ray-by-ray approximation.  Both of our simulations were performed on an axis-free Yin-Yang grid~\cite{2004GGG.....5.9005K,Wongwathanarat:2010yi}. 

In hydrodynamical simulations of BH-forming  models, where the central density and mass of the PNS continue to grow to relatively large values, an  implementation of general relativity (GR) is crucial. The VERTEX code approximates the effects of GR by replacing the Newtonian gravitational potential with a modified Tolman-Oppenheimer-Volkoff (TOV) potential, as proposed in~\cite{Rampp:2002bq,Marek:2005if} and implemented in~\cite{Pan:2017tpk, Hudepohl:2014,Mirizzi:2015eza,OConnor:2015rwy,Radice:2017ykv,Lentz:2015nxa,Kuroda:2018gqq}.  Specifically, by prescribing the TOV potential according to Case A in \cite{Marek:2005if}, the full GR results can be closely reproduced, while keeping a sophisticated treatment of neutrino transport and the simulation costs lower than needed for solutions of the Einstein field equations.  In particular, the use of the TOV potential allows for the reproduction of the duration of the time interval of accretion until BH formation and the corresponding neutrino energy release found in full GR 1D simulations fairly accurately~\cite{Hudepohl:2014,Mirizzi:2015eza,Schneider:2020kxr}. However, by adopting the approximate treatment of GR proposed in~\cite{Marek:2005if}, we have no way to recognize the formation of the 
event horizon, but we diagnose the onset of BH formation simply by a rapid rise of the central density and a corresponding drop of the lapse function. Redshift corrections and time dilation are included in the neutrino transport, while relativistic  transformations of the spatial coordinates are excluded in order to remain consistent with the Newtonian hydrodynamics equations governing the fluid dynamics in these simulations \cite{Rampp:2002bq}. 

The simulation of the $40\ M_\odot$ SN model has an angular resolution of $5$~degrees and a temporal resolution in the data output of $0.5$~ms. The simulation of the $75\ M_\odot$ SN model was run with the newly-implemented static mesh refinement grid~\cite{Melson:2019kjj}, i.e.~the neutrino transport was computed on a $2$-degree grid, but the hydrodynamic evolution was computed on a grid refined radially outwards up to $0.5$ degrees (doubling the angular resolution roughly at the position of the gain radius and at a fixed radius of $160$~km). The adopted temporal resolution of the $75\ M_\odot$ model is of about $0.2$~ms in the data output.  We stress that our angular resolution may not be completely satisfactory with respect to obtaining full convergence~\cite{Melson:2019kjj,Radice:2015qva,Nagakura:2019gmh}, however we accept this compromise for being able to choose high radial resolution with initially 400 radial zones, being non-equidistantly distributed from the center to an outer boundary at $10^4$~km with an inflow condition, and continuous refinement up to 650 radial zones towards the time when the simulations were stopped. Moreover, a comparison of the $75\ M_\odot$ model with a low-resolution run of $5$ degrees reveals essentially no differences in the dynamical evolution and neutrino emission (the main difference being a $4$~ms earlier approach to BH formation in the low-resolution case); for this reason, we discuss only the better resolved case here.  

The simulation of the $40\ M_\odot$ SN model employed the nuclear equation of state (EoS) of Lattimer and Swesty~\cite{Lattimer:1991nc} with nuclear incomprehensibility of $220$~MeV. The collapse into a BH occurs at $\sim 570$~ms post bounce. The simulation of the $75\ M_\odot$ SN model  again used the Lattimer and Swesty EoS, and the BH formation occurs at $\simeq 250$~ms post bounce, indicating a higher accretion rate and therefore a shorter accretion phase compared to the $40\ M_\odot$ model. Admittedly, the Lattimer and Swesty EoS adopted for both models is not consistent with all laboratory constraints of high-density nuclear matter~\cite{Oertel:2016bki}. More modern EoSs, for example those discussed in~\cite{Steiner:2012rk}, may yield a different maximum NS mass and therefore duration of post-bounce accretion until BH formation. Nevertheless,  we do not expect our general results and generic effects to be much affected by the shortcomings of the employed EoS. Accordingly, for example, a $5$-degree 3D simulation of the $75\ M_\odot$ progenitor with the SFHo EoS~\cite{Steiner:2012rk} exhibits a shock history very similar to the $75\ M_\odot$ run with the Lattimer and Swesty EoS, but BH-formation occurs only after $325$~ms, and a $75\ M_\odot$ simulation with a resolution of $5$ degrees with the DD2 EoS~\cite{Steiner:2012rk} shows a dynamical evolution very close to the $40\ M_\odot$ model with Lattimer and Swesty EoS until BH formation at $519$~ms after bounce. The results will be presented in detail in another work, and we  focus here on two representative cases to explore the imprints of BH formation carried by neutrinos.

\begin{figure*}
\centering
\includegraphics[width=0.85\columnwidth]{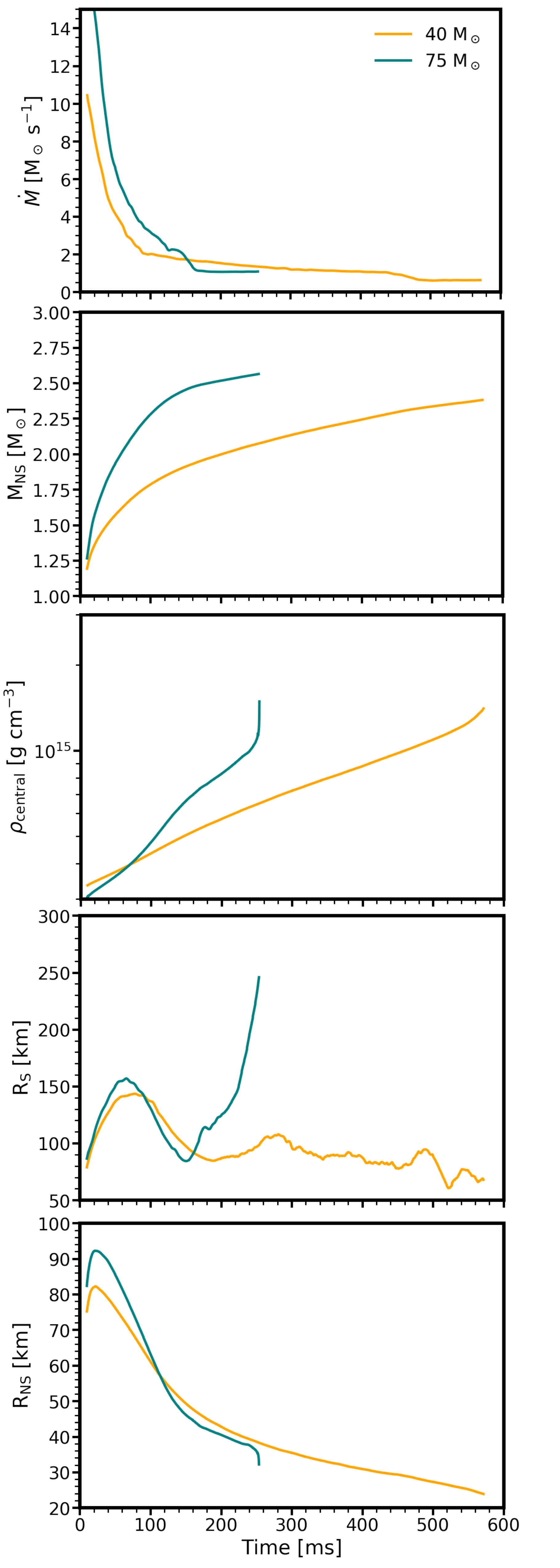}
\caption{Evolution of the pre-shock mass accretion rate (at 400 km),   NS baryonic mass, baryonic density at the NS center, spherically-averaged shock radius, and NS radius (defined at a density of $10^{11}$~g cm$^{-3}$) for the $40\ M_\odot$ (in orange) and $75\ M_\odot$ (in teal) models, from top to bottom. All curves have been smoothed with running averages over intervals of $5$~ms. The $40\ M_\odot$ model forms a BH at a post-bounce time of about $570$~ms, while the $75\ M_\odot$ model undergoes a more rapid evolution and forms a BH at a post-bounce time of roughly $250$~ms.  Shock expansion occurs in the $75\ M_\odot$ model  in the phase preceding the BH formation as the central density steeply increases, see also~\cite{Ott:2017kxl}. The shock radius exhibits a generally stationary value with slow evolution in the  $40\ M_\odot$ model. The $40\ M_\odot$ model shows episodes of shock contraction and expansion just before the BH formation. In both models, the NS baryonic mass rapidly increases, reaching a value of about $2.5\ M_\odot$ before the BH formation. }
\label{fig:simulations}
\end{figure*}

Figure~\ref{fig:simulations} displays the time evolution of the characteristic quantities for the two models. The $75\ M_\odot$ model settles to a generally constant  mass accretion rate after the infall of the  Si/O interface through the shock (top panel), and  a steep increase in the baryon density (third panel) in the instants preceding the BH formation. Due to stabilizing effects of  thermal pressure, BH formation sets in when a mass has been accreted onto the transiently stable PNS that  is  higher than the maximum mass of cold NSs for the employed EoS~\cite{Schneider:2020kxr}, which allows a maximum baryonic mass of about $2.3\ M_\odot$~\cite{Steiner:2012rk}. In both models, the PNS baryonic mass, plotted in the second panel of Fig.~\ref{fig:simulations}, rapidly increases and reaches a value of around $2.5\ M_\odot$ just before the BH formation.  

As can be seen in the plot of the shock radius evolution  (fourth panel),  shock expansion occurs in the $75\ M_\odot$ model just prior to the BH formation, similarly to what was observed in~\cite{Pan:2017tpk,Chan:2017tdg,Kuroda:2018gqq}. On the other hand,  this does not happen for the $40\ M_\odot$ model where the average shock radius reaches a quasi-stationary value, although with considerable excursions with time.  This trend in the shock radius for the $40\ M_\odot$ model differs from that of the $40\ M_\odot$ model simulated in 2D with the same EoS and nuclear incompressibility in~\cite{Pan:2017tpk}. In fact, the latter showed a shock revival prior to the collapse of the PNS into a BH. Similarly, the simulation of a $40\ M_\odot$ model considered in~\cite{Ott:2017kxl}, albeit with a different EoS, exhibits a shock expansion at a post-bounce time of $\simeq 200$~ms. Figure~\ref{fig:simulations} also shows that, in our $40\ M_\odot$ model, episodes of shock contraction and expansion occur just before the BH formation, while in our $75\ M_\odot$ model, the shock radius quickly expands before the PNS collapses into a BH.  As expected and shown in the bottom panel of Fig.~\ref{fig:simulations}, the PNS radius  rapidly contracts until the BH forms.

\section{Neutrino Emission Properties}\label{sec:Properties}

The neutrino signal carries unique imprints of the hydrodynamical instabilities governing the accretion phase. To determine those characteristic of the onset of BH formation, we explore the evolution of the emitted neutrino properties of our $40\ M_\odot$ and $75\ M_\odot$ models. 
The neutrino properties were extracted at a radius of $500$~km and remapped from the Yin-Yang simulation grid onto a standard spherical grid. The neutrino emission properties for each flavor have  been projected and plotted to appear as they would for a distant observer located along a specific angular direction, following the procedure outlined in Appendix A of ~\cite{Tamborra:2014hga}  and in~\cite{Muller:2011yi}. We emphasize that the observer projections are a post-processing measure intended to help overcome the shortcomings of the ray-by-ray transport (i.e., the fact that the ray-by-ray approximation considers only radial fluxes, whereas distant observers receive neutrinos from all locations on the hemisphere facing the observer). As a consequence, small local variations in the neutrino emission properties connected to the ray-by-ray transport approximation will be hemispherically averaged out through the observer projections. The neutrino emission properties for every angular direction, as well as the $4 \pi$-equivalent ones, can be provided upon request for \href{https://wwwmpa.mpa-garching.mpg.de/ccsnarchive/data/Walk2019/index.html}{both models}.

\subsection{Directionally independent neutrino emission properties}\label{sec:Other_Features}

Before investigating the directional dependence of the neutrino emission properties, we focus on characterizing the global features of the $40\ M_\odot$ and $75\ M_\odot$ models that may be unique to BH-forming stellar collapses. We do this by considering the neutrino emission properties obtained after integrating over all observer directions. 

\begin{figure*}
\centering
\includegraphics[width=1.7\columnwidth]{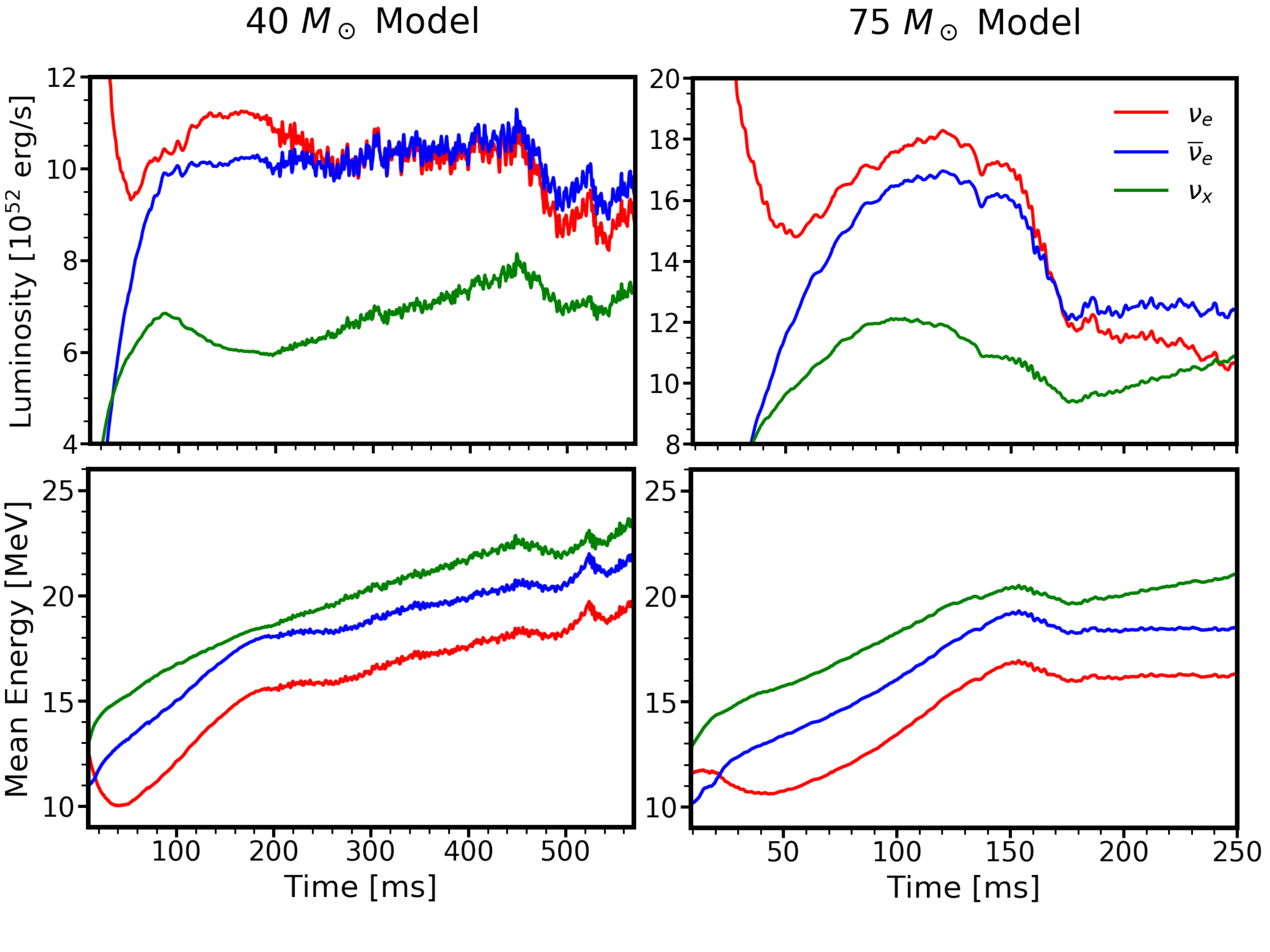}
\caption{{\it Top panels:} Luminosity of each neutrino flavor ($\nu_e, \overline{\nu}_e$ and  $\nu_x$ in red, blue, and green, respectively) for the $40\ M_\odot$ (left) and $75\ M_\odot$  (right) models after integrating over all observer directions. The luminosity for the $40\ M_\odot$ model  drops at $\sim 450$~ms corresponding to the infall of the Si/O interface and at $\sim 510$~ms, tracking the shock contraction and expansion just before the BH formation. Notably,  SASI modulations in the neutrino luminosity are still clearly visible in the $40\ M_\odot$ model, hinting towards very strong SASI activity.  A steady increase in the $\nu_x$ luminosity is observed for both models in more or less  the same temporal interval where  SASI occurs, as well as a crossing between the $\nu_e$ and $\bar{\nu}_e$ luminosities just before the onset of BH formation. {\it Bottom panels:} Mean energies (i.e., energy flux divided by number flux) of each neutrino flavor as a function of time. As opposed to the luminosity, the average energies show the same hierarchy throughout the whole signals duration. The luminosities and mean energies have been evaluated at a radius of 500 km.}   
\label{fig:Luminosity_4pi}
\end{figure*}

Both of our models develop long-lasting phases of SASI activity. Spiral SASI motions of the post-shock layer partly continue for periods of more than $100$~ms and can change their direction and main plane of the spiral motions. In addition, there are also short episodes of SASI dipolar and quadrupolar sloshing activity that precede the SASI spiral phases and bridge phases where the spiral motions possess different direction vectors.

The top panels of Fig.~\ref{fig:Luminosity_4pi} show the luminosity of each neutrino flavor ($\nu_e$, $\bar{\nu}_e$, and $\nu_x  = \bar{\nu}_x = \nu_\mu, \nu_\tau$) extracted at 500~km and individually obtained by integrating over all observer directions for both models. The red, blue, and green curves refer to $\nu_e$, $\overline{\nu}_e$, and $\nu_x$,  respectively. One can see that the large amplitude modulations due to  spiral SASI~\cite{Blondin:2006yw,Yamasaki:2007dc,Blondin:2006fx} survive in the $40\ M_\odot$ model (left panel), even after averaging over all observer directions, suggesting that spiral SASI is quite strong in this model. This is in agreement with the findings of Ref.~\cite{Vartanyan:2019ssu}, which also found strong spiral SASI activity in the context of compact shocks. On the contrary, this is not the case for the $75\ M_\odot$ model (right panel). Moreover, the $75\ M_\odot$ model has a shorter accretion phase prior BH formation and a higher overall neutrino luminosity. 

Comparing the global neutrino properties of these models with the ones of ordinary SNe, see e.g.~the neutrino properties of the SN models presented in~\cite{Tamborra:2014hga,Walk:2019ier,Walk:2018gaw} simulated with the same neutrino transport and nuclear EoS and the large suite of models in~\cite{Vartanyan:2019ssu}, we find that the neutrino luminosity is higher for the BH-forming models, up to a factor of two for the $75\ M_\odot$ model. The neutrino signal  terminates at $572$ and $250$~ms for the $40\ M_\odot$ and $75\ M_\odot$ models, respectively, corresponding to BH formation~\cite{1988ApJ...334..891B}. After $\sim 140-160$~ms post bounce, the $\nu_e$  luminosity---and after $\sim 100$~ms the $\nu_x$ luminosity---drop moderately in the $40\ M_\odot$ model because of the constant mass accretion rate but rapid PNS contraction (see top and bottom panels of Fig.~\ref{fig:simulations}). This decline is on a minor scale compared to the effect of the Si/O shell interface on the neutrino luminosity in the $40\ M_\odot$ model at around $440$--$480$~ms, and it happens  more gradually over a longer period of time from about $100$ to $220$~ms. Since the mass accretion rate ahead of the shock, during this time interval of the luminosity decline, is nearly constant and  the average shock even retreats (enhancing the PNS accretion until $\sim 190$~ms), we have no other explanation for this phenomenon than the ongoing, rather pronounced, contraction of the PNS, leading to a decrease of the neutrinospheric radius. After $\sim 190$~ms post bounce, shock expansion sets in and supports the decline of the $\nu_e$ luminosity by reducing PNS accretion. After about $260$--$270$~ms post bounce, a transient phase of slowly rising luminosities of all neutrino species follows.  At $\sim450$~ms post bounce, the infall of the Si/O interface in the $40\ M_\odot$ model leads to another temporary decline in the neutrino luminosity and to shock expansion.  Notably, the neutrino luminosity increases  after this drop, until $\sim 510$~ms, when it drops again. After this second drop, the luminosity climbs until the onset of the collapse to a BH.  This late behaviour tracks the contraction and expansion of the shock radius observed in Fig.~\ref{fig:simulations}. Around $\sim 550$~ms, the shock radius contracts again (see fourth panel of Fig.~\ref{fig:simulations}), leading to an increase in the luminosity and finally to the onset of BH formation.

Another notable feature, present in both BH-forming models, is the steady increase in the $\nu_x$ luminosity in  $[150,400]$~ms for the  $40\ M_\odot$ model  and after the drop due to the crossing of the Si/O interface at $175$~ms for the $75\ M_\odot$ model. This trend in the $\nu_x$ luminosity is characteristic of BH-forming models with high mass accretion rate. It may be explained by a significant increase in temperature as the PNS contracts, similar to what was found in~\cite{Liebendoerfer:2002xn}. The  higher temperatures lead to an increased production of $\mu$- and $\tau$-neutrinos. Moreover, a crossing between the $\nu_e$ and $\bar\nu_e$ luminosities occurs. In fact, the $\nu_e$ luminosity tends to steadily drop after the crossing of the Si/O interface at $450$~ms ($175$~ms) for the $40\ M_\odot$ ($75\ M_\odot$), while the $\bar\nu_e$ luminosity stays almost stationary.  This decrease of the $\nu_e$ luminosity is more pronounced in the simulation of the $75\ M_\odot$ model, which  also has a more massive and more rapidly contracting PNS, suggesting a possible connection between the two. Although the crossings of the electron neutrino and antineutrino luminosities are an interesting feature appearing in these models, they are not in the focus of our present discussion;  additional analysis and more models with different EoSs are needed in 1D, 2D and 3D to clarify the exact origin of these crossings and to reveal whether they are generic to BH formation in 3D.  The mean energies in the lower panels of  Fig.~\ref{fig:Luminosity_4pi} are calculated   by dividing the energy flux by the number flux. No crossing of the mean energies of $\bar\nu_e$ and $\nu_x$ occurs, in contrast to what was found in core-collapse simulations with lower mass accretion rates and correspondingly less rapidly growing PNS masses (see e.g.~\cite{Marek:2008qi}). Notably, the rise of the mean energy of $\nu_x$ is milder in these models than observed in 1D simulations of BH-forming stellar collapses by other groups~\cite{Fischer:2008rh,Sumiyoshi:2008zw} because of the inclusion of energy transfer in neutrino-nucleon scattering reactions in the neutrino transport of our models (for the correspondingly similar behavior in 1D models, see~\cite{Mirizzi:2015eza}).

\subsection{Directionally dependent neutrino emission properties}\label{sec:Directional_SASI}

Previous work~\cite{Tamborra:2014hga,Tamborra:2013laa,Walk:2018gaw,Walk:2019ier,Kuroda:2017trn,Takiwaki:2017tpe,Takiwaki:2016qgc,Lund:2012vm,Lund:2010kh,Vartanyan:2019ssu,OConnor:2018tuw} pointed out that the neutrino emission properties are highly directionally dependent. More specifically, in SN models showing SASI activity, sinusoidal modulations in the neutrino signal   are associated with periodic deformations of the shock wave. In the case of sloshing or spiral SASI, such modulations  are expected to be visible to an observer along the SASI direction or SASI plane, while they may nearly disappear for an observer located perpendicular to the SASI plane~\cite{Tamborra:2013laa, Tamborra:2014hga,Vartanyan:2019ssu}. Building on these findings, we attempt to identify potential  SASI episode(s) in the BH-forming models by looking for modulations in the neutrino signal and scanning over all observer directions.

\begin{figure*}
\centering
\includegraphics[width=2.\columnwidth]{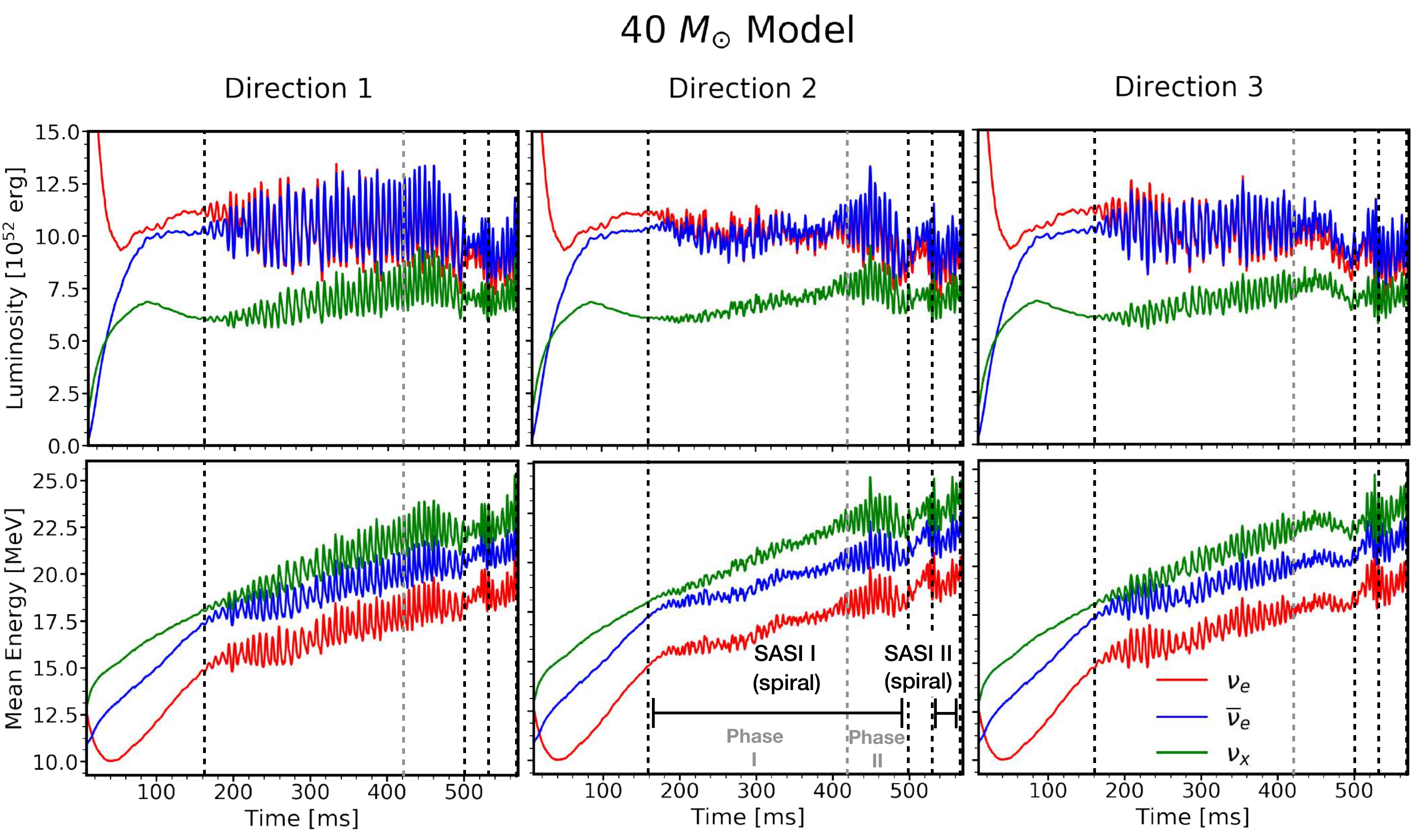}
\caption{Luminosities (top) and mean energies (bottom) of $\nu_e, \overline{\nu}_e$ and $\nu_x$ (red, blue, and green, respectively), along three observer directions for the $40\ M_{\odot}$ model as  functions of  post-bounce time. The motivation for selecting the three observed directions shown in this plot will be discussed in Sec.~\ref{sec:Evolution_SASI}, see also  Fig.~\ref{fig:s40_Lum_Dip_Evo}.  A long-lasting spiral SASI phase (SASI I) is identified along Direction 1 (left panel) in  $[160, 500]$~ms, as indicated by the black  dashed vertical lines. By comparing the neutrino signal along Directions 1 and 2, this interval can be divided into two sub-intervals, $[160, 420]$~ms  and $[420, 500]$~ms, divided by the grey dashed line. In fact, the SASI plane evolves over time, shifting its orientation at $\sim 420$~ms. Modulations corresponding to a second spiral SASI episode (SASI II) are visible in the interval $[530, 570]$~ms.  The spiral nature of the SASI episodes is further investigated in Sec.~\ref{sec:Evolution_SASI}.}
\label{fig:s40_Properties}
\end{figure*}

Figure~\ref{fig:s40_Properties} shows the neutrino luminosities and mean energies for the $40\ M_\odot$ model as a function of the post-bounce time  along three selected observer directions. The three directions are specifically chosen to highlight the most extreme modulation amplitudes of the neutrino signal, and to show the maximal variation between periods of modulation.  The precise choice of the location of the three selected directions relative to the orientation of the spiral SASI dipole will be discussed in Sec.~\ref{sec:Evolution_SASI}.

Along Direction 1, a period of high amplitude modulation of the neutrino properties can be observed in the interval $[160, 500]$~ms. These modulations are  indicative of a single long spiral SASI phase along the plane of observation  (SASI I). By comparing the left and central panels of Fig.~\ref{fig:s40_Properties}, however, one can see that only the modulations in  $[160, 420]$~ms decrease upon changing to an observer along Direction 2 (SASI I, Phase I). The modulations in the second sub-interval, $[420,  500]$~ms, disappear by shifting to an observer placed along Direction 3 (SASI I, Phase II). This suggests a change of the main spiral SASI plane at the  interval $[420, 500]$~ms. Whether the shift of the spiral SASI plane occurs gradually or instantaneously cannot be easily inferred by scanning through the observer directions alone, and will be further investigated in the next Section. Finally, a second period of signal modulation can be identified in Fig.~\ref{fig:s40_Properties} between $[530, 570]$~ms (SASI II). This indicates the occurrence of a second spiral SASI episode for the $40\ M_\odot$ model, which develops directly after the infall of the Si/O interface, and appears to remain stable until the collapse into a BH. 

\begin{figure*}
\centering
\includegraphics[width=1.5\columnwidth]{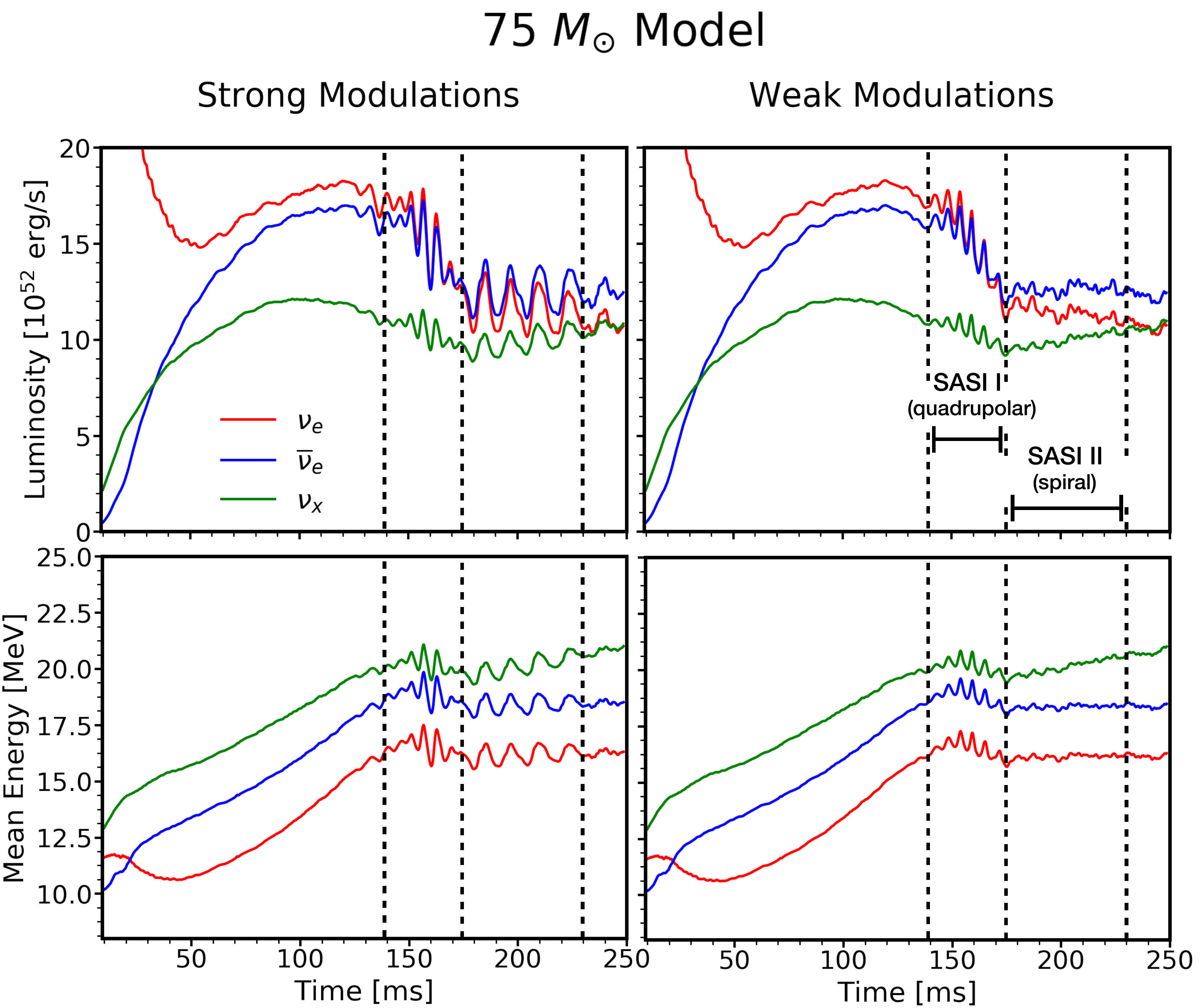}
\caption{Analogous to Fig.~\ref{fig:s40_Properties}, but for the $75\ M_\odot$ model. For this model, only two directions, the one of strongest (left) and weakest (right) overall signal modulation, are displayed to identify the spiral SASI phases, as will be further discussed in Sec.~\ref{sec:Evolution_SASI}.  The coordinates of the observer corresponding to these two directions are given in the lower two panels in Fig.~\ref{fig:s40_Lum_Dip_Evo}.  A quadrupolar moment dominates the neutrino emission  between $[140, 175]$~ms. This is followed by  dipolar modulations of the neutrino emission due to spiral SASI in $[175, 230]$~ms. Note that the quadrupolar modulations are due to a quadrupolar SASI mode and do not depend strongly on the direction.  Similar to the $40\ M_\odot$ model, the second SASI episode has a spiral nature, which is investigated in Sec.~\ref{sec:Evolution_SASI}.}
\label{fig:u75_Properties}
\end{figure*}

Similarly, Fig.~\ref{fig:u75_Properties} shows the neutrino properties of each flavor for the $75\ M_\odot$ model. Upon scanning over all observer directions, two directions are extracted to  illustrate the behavior of this model; a direction of ``weak modulations," in which there is minimal total variation of the signal compared to the average variation, and a direction of ``strong modulations," where the amount of signal variation is maximal. A first phase of signal modulations is apparent in the interval between $[140, 175]$~ms. These modulations remain present in the signal throughout all scanned observer directions and are due to SASI quadrupolar sloshing motions, as will be evidenced and investigated in the next Section. Following this, a spiral SASI phase can be identified in the interval of $[175, 230]$~ms. As expected, the latter depends more strongly on the observer direction, thus disappearing when shifting between the directions of strong and weak modulations. 
 
For both models, we find that the SASI modulations affecting the neutrino signal are stronger for $\nu_e$ and $\bar\nu_e$  than for $\nu_x$. Similar results were also found in Refs.~\cite{Tamborra:2014hga,Walk:2019ier,Vartanyan:2019ssu}.

\section{Characterization of SASI in black hole forming models}\label{sec:Evolution_SASI}

In order to obtain a better characterization of SASI in the BH-forming  models, and to verify the conjectures of the SASI episodes presented in the previous Section, a detailed investigation of the temporal evolution of SASI is required. This is the focus of this Section.

The full animation of the time evolution of the $\overline{\nu}_e$ luminosity relative to its $4\pi$-average, [$(L_{\bar{\nu}_e} - \langle L_{\bar{\nu}_e}\rangle)/\langle L_{\bar{\nu}_e}\rangle$] for the $40\ M_\odot$ model, provided as \href{https://wwwmpa.mpa-garching.mpg.de/ccsnarchive/data/Walk2019/index.html}{Supplemental Material} of this paper,  reflects that spiral SASI activity dominates the dynamics of this model. The development of a SASI spiral mode has been previously found in 3D hydrodynamical simulations, see e.g.~\cite{Blondin:2006yw,Vartanyan:2019ssu,Hanke:2013jat,Fernandez:2015yza,Fernandez:2010db,Summa:2017wxq,Kuroda:2016bjd}.

SASI motions of the post-shock layer modulate the accretion flow onto the PNS and thus lead to large-amplitude variations of the accretion luminosity~\cite{Tamborra:2014hga,Vartanyan:2019ssu}. As a consequence, a dominant dipole in the shock surface has been shown to be correlated with dominant dipole in the neutrino  luminosity, e.g.~see Fig.~3 of~\cite{Hanke:2013jat}, Fig.~8 of~\cite{Walk:2019ier}, Fig.~13 of~\cite{Summa:2017wxq}, and Fig.~18 of~\cite{Vartanyan:2019ssu}. Building on these findings and  following the approach introduced in Sec.~IV of \cite{Walk:2019ier}, we employ a multipole analysis of the shock surface and of the neutrino luminosity, and track the evolution of the multipoles in time. For simplicity, we limit our analysis to  the $\overline{\nu}_e$ signal; however, similar trends are found for the other (anti)neutrino species. 

\begin{figure*}
\centering
\includegraphics[width=1.5\columnwidth]{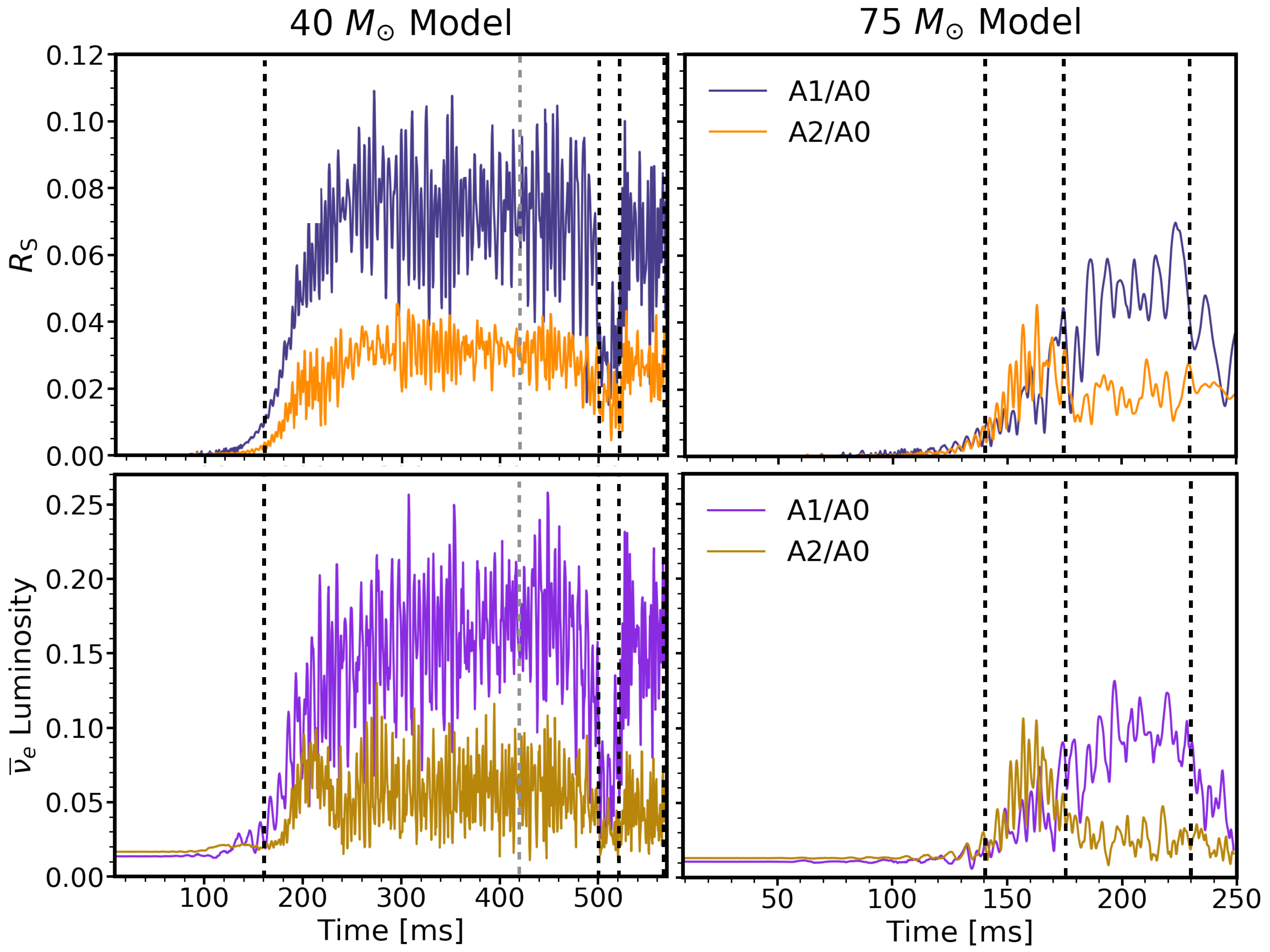}
\caption{Time evolution of the strength of the dipole (purple) and quadrupole (gold) moments of the shock surface (top) and $\overline{\nu}_e$ luminosity (bottom), each normalized to its monopole for the $40\ M_\odot$ (left) and $75\ M_\odot$ (right) BH-forming  models. The dashed vertical lines mark the spiral SASI intervals for each  model (see Sec.~\ref{sec:Directional_SASI}).  SASI dipole and quadrupole modes of the shock surface correlate tightly with the ones of the neutrino-emission asymmetries. The luminosity dipole dominates for both models throughout  the spiral SASI phases. For the $75\ M_\odot$ model, the quadrupole moment dominates in the interval $[140, 175]$~ms, confirming that the signal modulations observed in Fig.~\ref{fig:u75_Properties} (SASI I)  are due to a  quadrupolar mode in the neutrino emission.  The transition of a dominant quadrupolar ($l=2$) to a dominant dipolar ($l=1$) neutrino emission mode is a consequence of the rapid expansion of the shock radius caused by a  drop in the mass accretion rate.}
\label{fig:Rel_Lum_OBS}
\end{figure*}

We estimate the monopole ($A_0$), dipole ($A_1$), and quadrupole ($A_2$) as described in  Sec.~IV and Eq.~5 of~\cite{Walk:2019ier}. Figure~\ref{fig:Rel_Lum_OBS} shows that the spiral SASI dipole and sloshing quadrupole modes dominate in different phases and correlate tightly with the dominance of dipole or quadrupole neutrino-emission asymmetries. The left panels of Fig.~\ref{fig:Rel_Lum_OBS} show the time evolution of $A_1$ and $A_2$ relative to $A_0$ for the $\overline{\nu}_e$ luminosity and the shock surface of the $40\ M_\odot$  BH-forming  model. The dashed vertical lines indicate the two spiral SASI intervals described in Sec.~\ref{sec:Directional_SASI} (SASI I and SASI II). In both intervals, the dipole moment is dominant, confirming the development of spiral SASI in each time window, compatibly with what was discussed in Refs.~\cite{Vartanyan:2019ssu,Walk:2019ier}.

\begin{figure*}
\centering
\includegraphics[width=2\columnwidth]{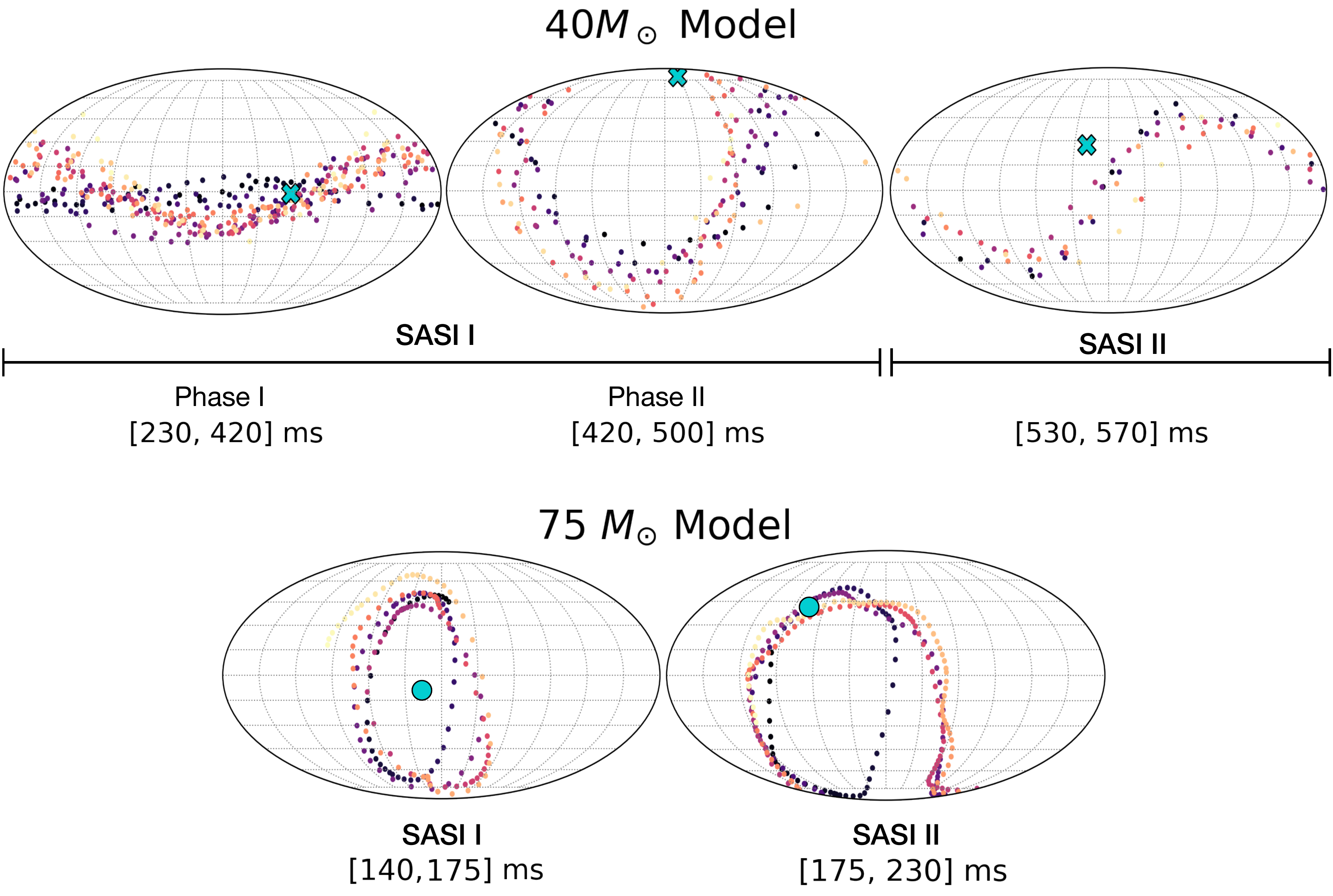}\\
\caption{Time evolution of the direction of the $\overline{\nu}_e$ luminosity dipole moment for the different  SASI time intervals identified in Sec.~\ref{sec:Directional_SASI} for the $40\ M_\odot$ (top) and the $75\ M_\odot$ (bottom) models. The dots mark the path of the positive luminosity dipole direction, and the color hues become lighter as time increases. \textit{Top, left}:  The (SASI I, Phase I) episode has been shortened and plotted only in $[230, 420]$~ms to highlight its trend more clearly, as the luminosity dipole is less stable at earlier times. In this time interval, spiral  SASI motions develop roughly along the equatorial plane (see also Fig.~\ref{fig:Lum_Dip_Dir}). The cyan marker indicates the coordinates of an observer along Direction 1, see left column of Fig.~\ref{fig:s40_Properties}. \textit{Top, middle}: Same as the left panel but for  $[420, 500]$~ms (SASI I, Phase II). The cyan marker indicates the coordinates of an observer along Direction 2 (see Fig.~\ref{fig:s40_Properties}). Between post-bounce times of approximately $380$--$420$~ms, the SASI plane shifts to a plane almost perpendicular to the equator.  \textit{Top, right}: Track of the $\overline{\nu}_e$ luminosity dipole moment over the second spiral SASI episode (SASI II). The marker indicates the coordinates of an observer along Direction 3.  {\it Bottom}: Time evolution of the $\overline{\nu}_e$ luminosity dipole direction of the $75\ M_\odot$ model in the first ($[140, 175]$~ms, left) and second ($[175, 230]$~ms, right) SASI intervals (SASI I and SASI II, respectively;  see Fig.~\ref{fig:Lum_Dip_Dir}). The markers indicate the coordinates of an observer along the Strong Modulations direction (right) and the Weak Modulations direction (left), see Fig.~\ref{fig:u75_Properties}. In this model, the luminosity dipole follows a stable and quite narrow path.}
\label{fig:s40_Lum_Dip_Evo}
\end{figure*}

\begin{figure*}
\centering
\includegraphics[width=1.5\columnwidth]{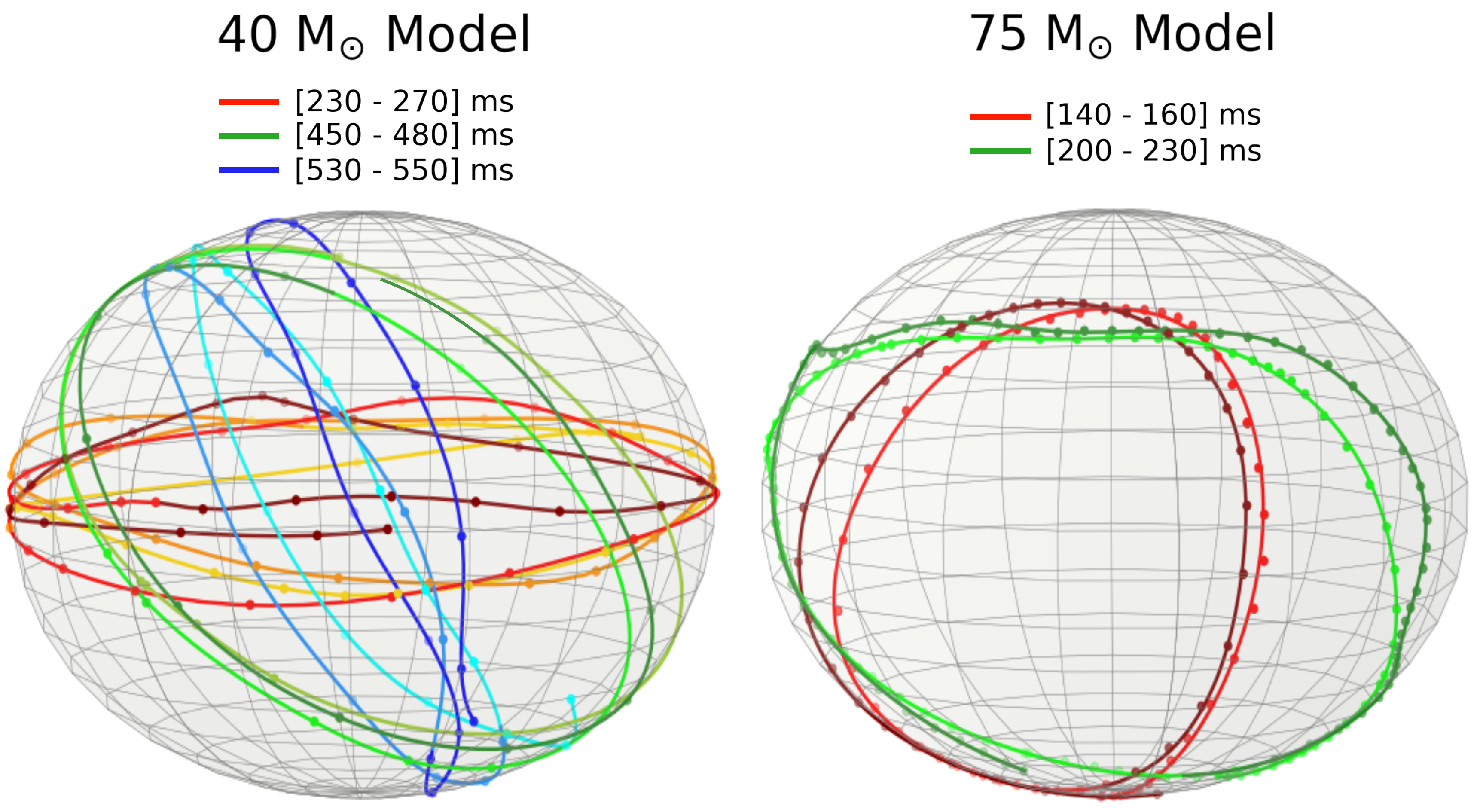}\\
\caption{Time evolution of the direction of the $\overline{\nu}_e$ luminosity dipole moment (see Figs.~\ref{fig:Rel_Lum_OBS} and \ref{fig:s40_Lum_Dip_Evo}) onto a 3D spherical map for selected short time intervals for the $40\ M_\odot$ (left) and the $75\ M_\odot$ (right) models. The red, green, and blue color codes represent different time intervals during the spiral SASI episodes of each model. Each individual full cycle of SASI is plotted with a different shade of color to indicate the time evolution. The dots mark the time evolution corresponding to  the temporal grid of the simulation output. The circular paths of the neutrino luminosity dipole clearly indicate the spiral nature of the SASI episodes in the $40\ M_\odot$ model. Similar conclusions hold for the second selected time interval (green) of the $75\ M_\odot$ model. During the earlier quadrupolar sloshing SASI phase of the $75\ M_\odot$ model (red), the dipole evolves along a $90^\circ$ slice in one hemisphere, while the quadrupolar moment dominates the neutrino emission (see also Fig.~\ref{fig:Rel_Lum_OBS}).}
\label{fig:Lum_Dip_Dir}
\end{figure*}

To investigate the shift of the spiral SASI plane at $\sim 420$~ms during the SASI I episode, the direction of the positive neutrino dipole moment is tracked along the SN emission surface over time in the top panels of Fig.~\ref{fig:s40_Lum_Dip_Evo}. The left and central panels show the first spiral SASI episode split up into the two sub-time intervals identified in Fig.~\ref{fig:s40_Properties} (SASI I, Phase I and II), and the right panel shows the second spiral SASI episode (SASI II).  The circular paths of the $\bar{\nu}_e$ dipole moment over the emission surface visible in Fig.~\ref{fig:s40_Lum_Dip_Evo} suggest the spiral nature of SASI in this model. To further highlight the spiral character of both spiral SASI episodes, the left panel of Fig.~\ref{fig:Lum_Dip_Dir} shows the $\bar\nu_e$ luminosity dipole evolution projected onto a 3D spherical map for short, representative time intervals.

From Fig.~\ref{fig:s40_Lum_Dip_Evo}, it is clear that the first spiral SASI phase kicks in along the equatorial plane.  Here, it remains relatively stable until $\sim 420$~ms. After that,  the spiral SASI  plane morphs from an equatorial one to the almost perpendicular plane spun by the circular pattern in the central panel of Fig.~\ref{fig:s40_Lum_Dip_Evo} (see also Fig.~\ref{fig:Lum_Dip_Dir}). Direction 1 (marked in the top left panel of Fig.~\ref{fig:s40_Lum_Dip_Evo}) lies on both the equatorial plane and this shifted plane, explaining the presence of signal modulations over the full interval, $[160, 500]$~ms, in the left panel of Fig.~\ref{fig:s40_Properties}. Direction 2 (see top middle panel of Fig.~\ref{fig:s40_Lum_Dip_Evo}), however, lies far away from the equator, and thus, the central panel of Fig.~\ref{fig:s40_Properties} does not show strong amplitude modulations in the first sub-time interval. The marker of Direction 3 (top right panel of Fig.~\ref{fig:s40_Lum_Dip_Evo})  lies away from the plane to which the second phase of the first spiral SASI episode evolves in the interval $[420, 500]$~ms (SASI II), and thus, the signal modulations in this interval shrink considerably between Direction 2 and Direction 3 in Fig.~\ref{fig:s40_Properties}.  The markers in Fig.~\ref{fig:s40_Lum_Dip_Evo} indicate the coordinates along which Directions 1, 2, and 3 in Fig.~\ref{fig:s40_Properties} lie, from left to right, respectively. When scanning over all observer directions, these directions were chosen because the total modulations along each of them differ maximally compared to the modulations visible in the other two directions.

The evolution of the spiral SASI plane over time is special for this BH-forming model, and may be observable due to the long duration of the first spiral SASI episode. SASI spiral modes are thought to develop as a result of a superposition of two SASI dipolar modes oscillating out of phase with each other along different directions~\cite{Blondin:2006yw,Fernandez:2010db,Hanke:2013jat,Vartanyan:2019ssu}. The shift of the plane of the SASI spiral mode observed in this model may be caused by the presence of a second, subdominant  SASI mode along a different direction or by a stochastic fluctuation. However, to fully understand the physical origin of the shift in the spiral SASI plane, one must consider the full range of hydrodynamical properties of this model. Such an investigation lies beyond the scope of this work, where the neutrino emission properties are the main focus.

\begin{figure*}
\centering
\includegraphics[width=1\columnwidth]{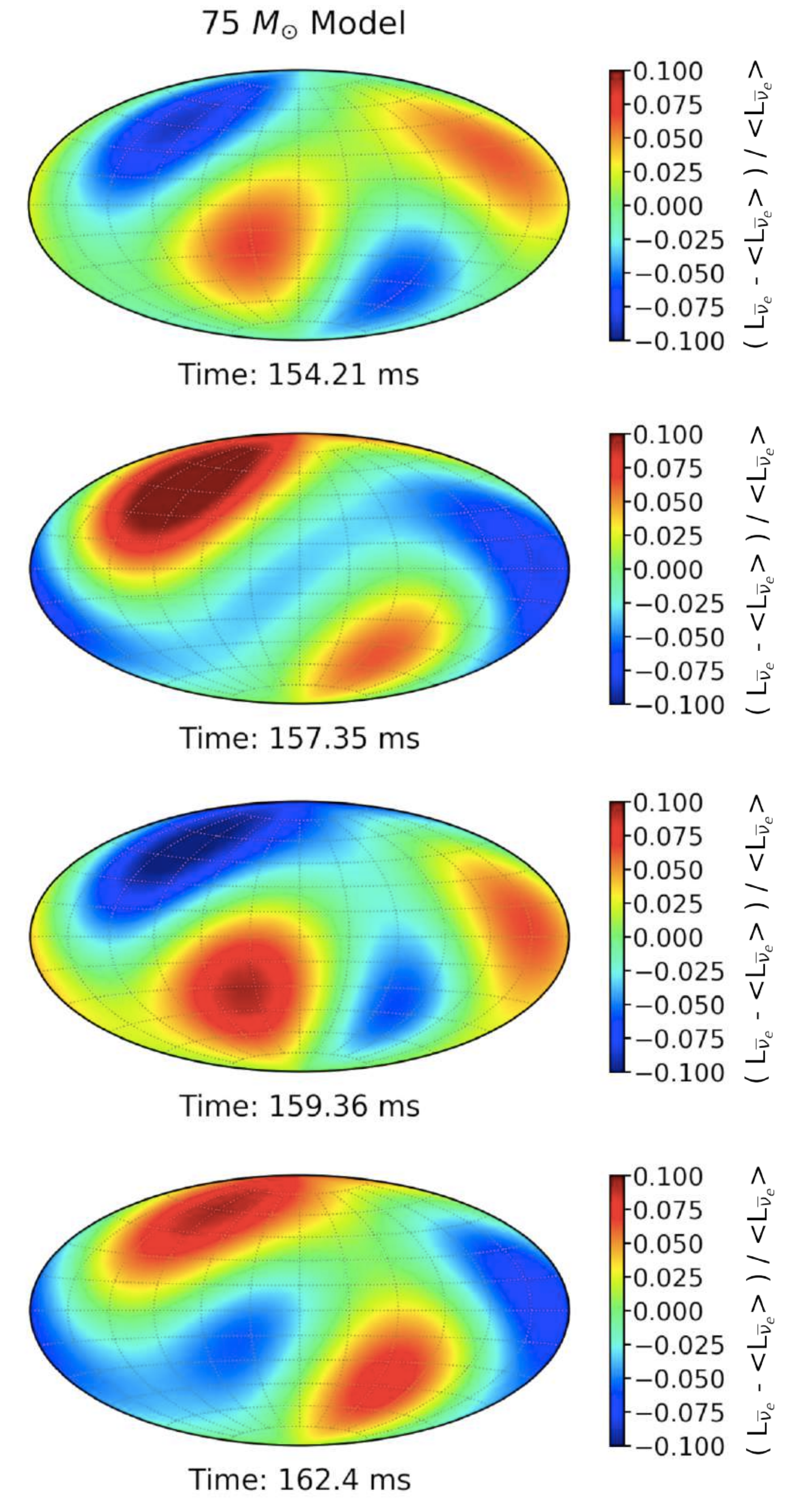}
\caption{Molleweide maps of  the $\overline{\nu}_e$ luminosity relative to its $4\pi$-average for the $75\ M_\odot$ model for four consecutive time snapshots in  $[140, 175]$~ms where the SASI quadrupolar  motions are dominant (see Fig.~\ref{fig:Rel_Lum_OBS}). The quadrupolar structure of the neutrino emission is clearly visible across the four  snapshots. This explains  the quasi-periodic, directionally independent signal modulations visible in Fig.~\ref{fig:u75_Properties}. } 
\label{fig:Quadrupole_Snapshots}
\end{figure*}

The right panels of Fig.~\ref{fig:Rel_Lum_OBS} show the corresponding relative dipole and quadrupole strengths of the $\overline{\nu}_e$ luminosity for the $75\ M_\odot$ model as  functions of time. Interestingly, in the right panels of Fig.~\ref{fig:Rel_Lum_OBS} the quadrupole moment  dominates in the earlier interval, $[140, 175]$~ms, confirming that the corresponding SASI modulations identified in the neutrino luminosity  have a quadrupolar sloshing nature.  This quadrupolar emission pattern correlates with a dominant quadrupolar sloshing mode of SASI mass motions of the post-shock layer as demonstrated by our spherical harmonics decomposition of the shock surface. In fact, by plotting the $\overline{\nu}_e$ luminosity relative to its $4\pi$-average ($\langle L_{\bar{\nu}_e}\rangle$) on a Molleweide map for  the SN emission surface at four consecutive snapshots in time in Fig.~\ref{fig:Quadrupole_Snapshots}, a clear quadrupolar pattern of SASI can be seen. The SASI quadrupolar sloshing activity  is even more apparent in the full animation of the $\overline{\nu}_e$ luminosity relative the $4\pi$-average, added as \href{https://wwwmpa.mpa-garching.mpg.de/ccsnarchive/data/Walk2019/index.html}{Supplemental Material} of this paper, from which the snapshots in Fig.~\ref{fig:Quadrupole_Snapshots} have been taken.  

Once the Si/O shell interface has fallen into the shock front at $\sim 175$~ms, the shock front expands, increasing the volume within the gain layer (see Fig.~\ref{fig:simulations}). This leads to favorable conditions for the development of a dipolar SASI mode~\cite{Foglizzo:2006fu} and thus of spiral SASI activity in the $75\ M_\odot$ model as shown in the right panels of Figs.~\ref{fig:Rel_Lum_OBS} and \ref{fig:Lum_Dip_Dir}, and it is consistent with the dominance of the spiral SASI mode in the $40\ M_\odot$ model, where the shock radius during most of the post-bounce evolution (until shortly before BH formation) is larger than in the $75\ M_\odot$ model between $140$~ms and $175$~ms.   The development of the SASI spiral mode is evidenced by the dominance of the  neutrino emission dipole over the emission quadrupole in $[175, 230]$~ms in the right panels of Fig.~\ref{fig:Rel_Lum_OBS}. This transition between an $l=2$ (quadrupolar) and an $l=1$ (spiral) SASI mode is triggered by the rapid shock expansion that is caused by the drop in the mass accretion rate, in agreement with the findings of~\cite{Foglizzo:2006fu,Scheck:2007gw,Hanke:2011jf,Guilet:2011aa,Fernandez:2013sqa,Ohnishi:2005cv}, and it is reflected in the neutrino signal, as also discussed in~\cite{Vartanyan:2019ssu}, where failed and successfull explosions were compared. 

The bottom panels of Fig.~\ref{fig:s40_Lum_Dip_Evo} display the evolution of the $\bar{\nu}_e$ luminosity dipole moment. The left panel shows that the $\bar{\nu}_e$ dipole direction oscillates only within a $90^\circ$ slice in one hemisphere during the SASI I phase of the $75\ M_\odot$ model in $[140,175]$~ms (see also the right panel of Fig.~\ref{fig:Lum_Dip_Dir}). This is compatible with the dominant quadrupolar SASI during this time interval of small shock radius. The bottom right panel of Fig.~\ref{fig:s40_Lum_Dip_Evo} depicts the evolution of the $\bar{\nu}_e$ luminosity dipole during the spiral SASI episode (SASI II), where the dipole vector describes a circular path around the whole sphere (see also right panel of Fig.~\ref{fig:Lum_Dip_Dir}). Clearly, the spiral SASI plane is stationary for the $75\ M_\odot$ model. The moment preceding BH formation is marked by a steep decay of the luminosity dipole at $\simeq 230$~ms. This, in turn, is also visible as a damping of the spiral SASI modulations in Fig.~\ref{fig:u75_Properties}. 

\section{LESA in black hole forming models}\label{sec:LESA}

 In \cite{Tamborra:2014aua} a large hemispheric dipolar asymmetry in the electron-lepton number (ELN) neutrino flux was found in the 3D simulations of $11.2$, $20$, and $27\ M_\odot$ SN models. These findings have been confirmed in~\cite{Janka:2016fox,Tamborra:2014hga} and, more recently,  in~\cite{OConnor:2018tuw,Vartanyan:2018iah,Glas:2018vcs,Vartanyan:2019ssu}. LESA is thought to originate from hemispherically asymmetric convection in the PNS, which induces regions of excess of $\nu_e$ relative to $\overline{\nu}_e$ in the PNS convective layer~\cite{Tamborra:2014aua,Glas:2018vcs}. This, in turn, leads to the development of large-scale asymmetries in the ELN flux and other medium-related properties, e.g.~a hemispheric asymmetry of the electron fraction in the PNS~\cite{Tamborra:2014aua}. On the other hand, also SASI motions of the post-shock layer can produce dipole and quadrupole modes in the neutrino emission as discussed in Sec~\ref{sec:Evolution_SASI}. In the following, we attempt to discriminate the LESA and SASI asymmetries in the neutrino signal. 
 
In order to quantitatively characterize the ELN emission asymmetries (caused by LESA or SASI),   we estimate the relative excess of $\nu_e$ over $\bar{\nu}_e$ emission for each angular direction $(\theta,\phi)$~\cite{Tamborra:2014aua}
\begin{equation}
\label{eq:sigma}
\Sigma = \frac{1}{T} \int_{t_1}^{t_2} \frac{N_{\nu_e}-N_{\bar\nu_e}}{\langle N_{\nu_e}+N_{\bar\nu_e} \rangle}\ ,
\end{equation} 
with $T=t_2-t_1$ being the time interval where the relative excess is estimated, $N_{\nu_e,\bar\nu_e}$  is the electron (anti)neutrino number flux, and $\langle N_{\nu_e,\bar\nu_e}\rangle$ the corresponding $4 \pi$-average of the neutrino number flux. 
 
\begin{figure*}
\centering
\includegraphics[width=2\columnwidth]{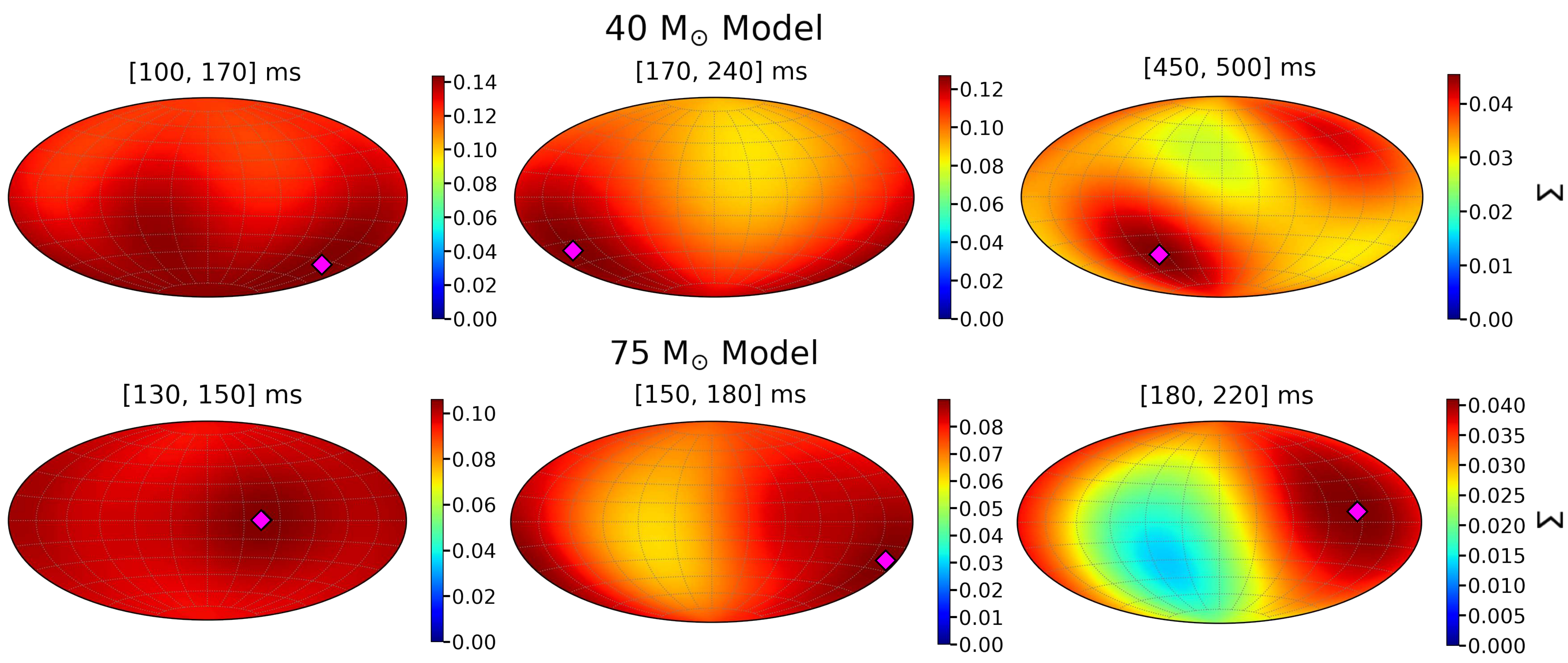}
\caption{Molleweide maps of the $\Sigma$ parameter (Eq.~\ref{eq:sigma}) for the $40\ M_\odot$ (top) and $75\ M_\odot$ (bottom)  models for three selected time intervals (from left to right, respectively). The  time intervals have been chosen to track the evolution of the  ELN dipole moment. The differences in magnitude of $\Sigma$  between hot and cold regions correspond to LESA activity because the time averages smooth out short-time variations due to  SASI emission modulations. $\Sigma$ is smaller in these BH-forming models in comparison to other core-collapse SN models previously analyzed (cfr., e.g., Fig.~6 of ~\cite{Tamborra:2014aua}). The $\Sigma$ parameter is largest in the earliest time interval (leftmost panels) which has been integrated for both models from $\mathcal{O}(100)$~ms post bounce until approximately the start of the first SASI episode for each model. The ``$\Diamond$'' marker indicates the location where $\Sigma$ is maximal in each selected time interval for each model.} 
\label{fig:Rho_Plot}
\end{figure*}

We use Eq.~\ref{eq:sigma} to estimate the strength of the long-time stable ELN emission asymmetries in our BH-forming models, and for that average over sufficiently long time intervals to smooth out the short-timescale  SASI variations. The results are shown in Fig.~\ref{fig:Rho_Plot}. To track any effects due to the presence of stable (in time and space) ELN emission asymmetries, which are likely to be caused by LESA and not  SASI, we choose to integrate over three time intervals for each model (i.e., $[100,170]$~ms, $[170,240]$~ms, and $[450,500]$~ms  for the $40\ M_\odot$ model, and $[130,150]$~ms, $[150,180]$~ms, and $[180,220]$~ms for the $75\ M_\odot$ model). Both models show a significantly smaller $\Sigma$  than was previously found in standard core-collapse SN models (see, e.g., Fig.~6 of~\cite{Tamborra:2014aua} and Refs.~\cite{Walk:2019ier,OConnor:2018tuw,Vartanyan:2018iah,Vartanyan:2019ssu,Tamborra:2014hga} for comparison). The strength of $\Sigma$ is greatest in the earliest time interval. The weak nature of  the emission asymmetries relative to the monopole or direction-averaged electron neutrino and antineutrino emission  in these BH-forming models is caused by the very  high mass accretion rate of the PNS before collapse to a BH. The neutrino emission is strongly dominated by the  SASI-modulated accretion luminosity.

\begin{figure*}
\centering
\includegraphics[width=2\columnwidth]{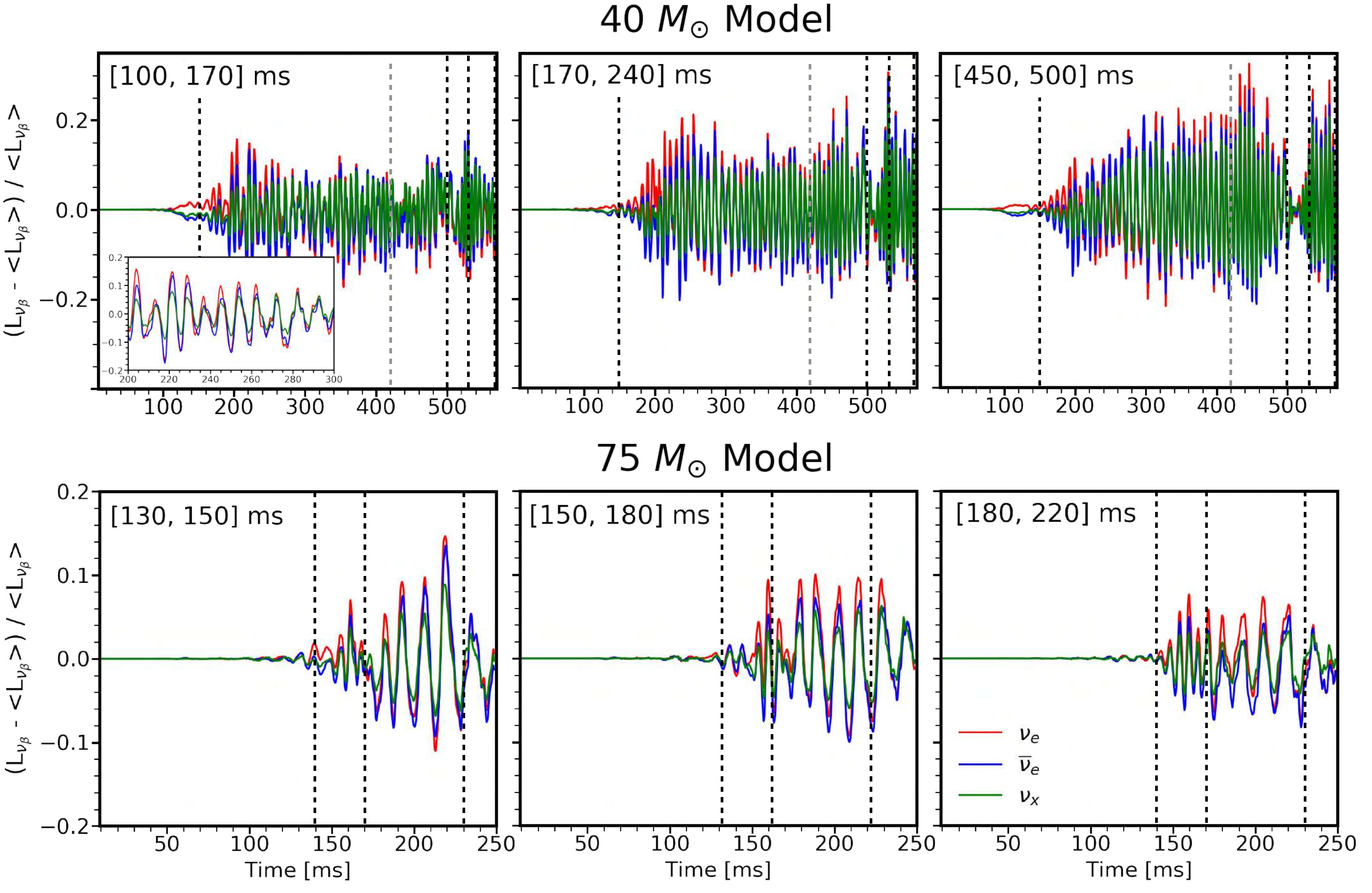}
\caption{Luminosity evolution of each neutrino flavor ($L_{\nu_\beta}$) for the $40\ M_\odot$ (top) and $75\ M_\odot$ (bottom) models along the ``$\Diamond$'' directions (selected for each time interval as indicated in Fig.~\ref{fig:Rho_Plot}) relative to the time-dependent average luminosity over all directions ($\langle L_{\nu_\beta} \rangle$). The dashed lines mark the  SASI intervals. Because of LESA, the  luminosity variations  of $\nu_e$ and $\bar{\nu}_e$ show a small anti-correlated displacement. The inset in the top left panel highlights only a minimal anti-correlated displacement of the $\nu_e$ signal to the positive side and of the $\bar{\nu}_e$ signal to the negative side, because at later times the ``$\Diamond$'' direction moves to different locations. The shift of the $\nu_e$ relative to the $\bar\nu_e$ luminosity is much smaller than in Fig.~5 of~\cite{Tamborra:2014aua} and Figs.~4, 7, 8, 10, 11 of~\cite{Tamborra:2014hga}. These findings suggest that the LESA activity affects the neutrino emission on a very low level. }
\label{fig:Rel_Lum}
\end{figure*}

References~\cite{Tamborra:2014aua,Tamborra:2014hga} found that the development of LESA is characterized by an anti-correlation of the relative luminosities of $\nu_e$ and $\bar{\nu}_e$ (cf.~Fig.~5 of~\cite{Tamborra:2014aua} and Figs.~4, 7, 8, 10 and 11 of~\cite{Tamborra:2014hga}), while SASI induces fully correlated variations of the $\nu_e$ and $\bar{\nu}_e$ luminosities and also a possibly associated ELN dipole. Figure~\ref{fig:Rel_Lum} shows the evolution of the  luminosity of each neutrino flavor ($L_{\nu_\beta}$) relative to the time-dependent average luminosity over all directions ($\langle L_{\nu_\beta} \rangle$).  The selected observer directions correspond to the coordinates of the ``$\Diamond$" directions shown in Fig.~\ref{fig:Rho_Plot} for the $40\ M_\odot$ and $75\ M_\odot$ models. The ``$\Diamond$" direction is the direction along which $\Sigma$ is maximal for each model in each selected time interval, and thus we expect to see the maximum anti-correlations in the neutrino vs.~antineutrino emission properties along this direction. Figure~\ref{fig:Rel_Lum} shows small anti-correlated displacements of the  luminosity variations of $\nu_e$ and $\bar{\nu}_e$ outside the SASI intervals, slightly visible for the $40\ M_\odot$ model until $\sim 200$~ms and tiny in the $75\ M_\odot$ model.  During the SASI-modulated episodes, a low-level displacement of the $\nu_e$ luminosity relative to the $\bar{\nu}_e$ one can be seen (see the panels of Fig.~\ref{fig:Rel_Lum} that show the ``$\Diamond$'' directions of Fig.~\ref{fig:Rho_Plot} during the second and third time intervals). The displacements of $\nu_e$ and $\bar\nu_e$ peaks in opposite directions are much smaller than those found in~\cite{Tamborra:2014aua,Tamborra:2014hga}. This points to a negligible impact of  LESA on the neutrino emission properties in both BH-forming  models, much smaller than seen in the previous models.  Notably, both BH-forming models are affected by the very strong spiral SASI activity that instead drives the $\nu_e$ and $\bar{\nu}_e$ luminosities in phase with each other and rules over the effects of the weak LESA. This sub-dominance of  LESA compared to SASI in the relative luminosity is clearly visible in the inset of Fig.~\ref{fig:Rel_Lum}. 

\begin{figure*}
\centering
\includegraphics[width=1.5\columnwidth]{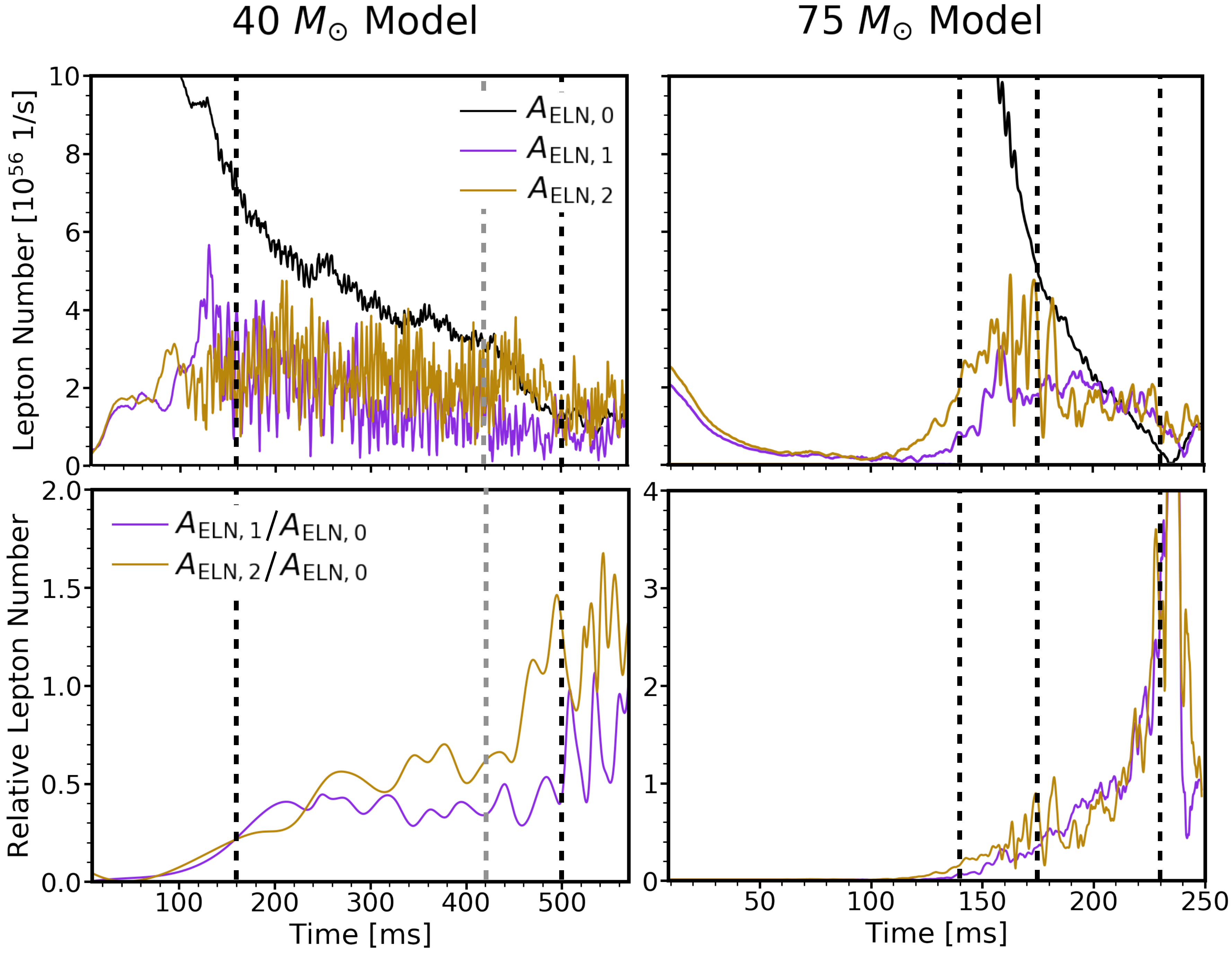}
\caption{Time evolution of the amplitudes of the ELN multipole moments for the $40\ M_\odot$ (left) and $75\ M_\odot$ (right) BH-forming models. The dashed lines mark the SASI intervals. {\it Top:} Evolution of the ELN monopole $l=0$ (in black), the dipole $l=1$  (purple) and the quadrupole $l=2$ (gold). {\it Bottom:} Evolution of the ELN dipole (purple) and quadrupole (gold) relative to the monopole (in the bottom left panel, the curve has been smoothed to highlight the temporal evolution despite the high-frequency fluctuations). The ELN dipole and quadrupole are comparable in absolute values to the ones previously found in core-collapse SN models (see e.g.~Fig.~3 of~\cite{Tamborra:2014aua} and Fig.~10 of~\cite{Walk:2019ier} for a direct comparison). However, the monopole is larger, and the dipole and quadrupole of the ELN emission are  mainly driven by the strong (short-time modulated) spiral SASI activity responsible for an ELN accretion dipole (quadrupole) in these BH-forming models (see Fig.~\ref{fig:Rel_Lum} for comparison).}
\label{fig:LESA_multipoles}
\end{figure*}

Both the LESA phenomenon and SASI-modulated accretion are responsible for inducing a dipole (quadrupole) in the ELN emission~\cite{Tamborra:2014hga}. In order to quantify the time evolution of the ELN multipoles and decipher any features due to LESA, we adopt a decomposition in spherical harmonics of the ELN emission following  Sec.~IV B of \cite{Walk:2019ier}. Figure~\ref{fig:LESA_multipoles}  shows the time evolution of the ELN monopole, dipole and quadrupole ($A_{\mathrm{ELN},0}, A_{\mathrm{ELN},1}$, and $A_{\mathrm{ELN},2}$, respectively). For both models, the dipole and quadrupole are comparable in amplitude to those of other core-collapse SN models, see e.g.~Fig.~3 of~\cite{Tamborra:2014aua} and Fig.~10 of~\cite{Walk:2019ier} for a direct comparison among SN models from this group as well as~\cite{OConnor:2018tuw,Vartanyan:2019ssu,Vartanyan:2018iah}. For the $40\ M_\odot$ model, a growing dipole (quadrupole) is observable before $\sim 150$~ms (in correspondence with the LESA anti-correlation between the $\nu_e$ and $\bar\nu_e$ relative luminosities visible in Fig.~\ref{fig:Rel_Lum}),  but the monopole is much larger. As the SASI I and SASI II phases kick in, a short-time variable dipole (quadrupole) is observable. For the $75\ M_\odot$ model,  the ELN quadrupole (dipole) appears  mostly due to the SASI quadrupole  (dipole); in fact,  the ELN quadrupole (dipole) in Fig.~\ref{fig:LESA_multipoles} grows in the same time interval when the $\nu_e$ and $\bar\nu_e$ relative luminosities are almost fully correlated in Fig.~\ref{fig:Rel_Lum}, and the LESA anti-correlation between the $\nu_e$ and $\bar\nu_e$ relative luminosities is minimally visible  in Fig.~\ref{fig:Rel_Lum}. Our findings lead us to conclude that, in the present BH-forming models, the accretion luminosity and its SASI modulations largely dominate the neutrino emission; as a consequence,  the SASI dipole and quadrupole are mainly visible, but hardly or not at all the LESA multipoles. As we will see in the following, also other signatures of LESA are only weakly developed in these BH-forming models.

\begin{figure*}
\centering
\includegraphics[width=1.8\columnwidth]{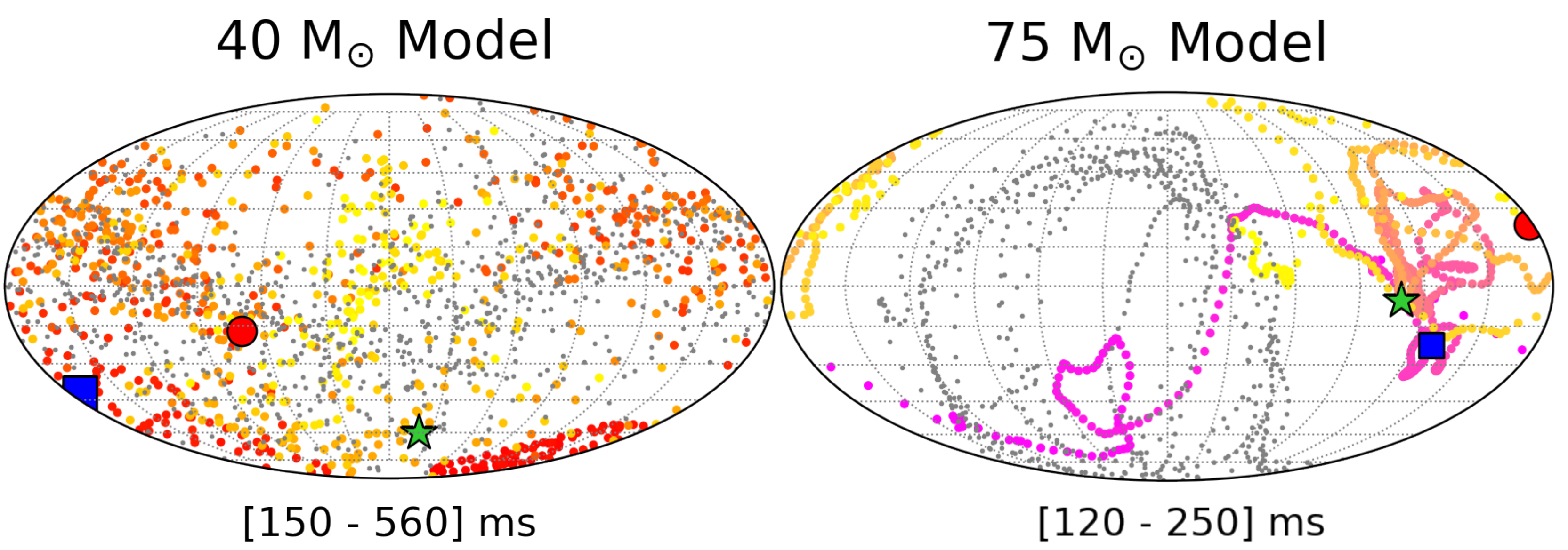}
\caption{Time evolution of the direction of the ELN dipole moment for the $40\ M_\odot$ (left) and $75\ M_\odot$ (right) models during the time intervals when the ELN dipole is dominant or equally strong as the quadrupole. The colorful dots mark the path of the positive ELN dipole direction and the color hues become lighter as time increases. The grey dots  mark the path  of the positive $\bar\nu_e$ luminosity dipole moment following the spiral SASI evolution (see Fig.~\ref{fig:s40_Lum_Dip_Evo}). In each map, the markers indicate the directions of the positive ELN dipole at the three time snapshots depicted in Fig.~\ref{fig:Ye}. The ELN dipole does not reach  stationary conditions for the $40\ M_\odot$ model, but instead rapidly wanders across the whole emission surface. The dipole migrates only within one half of the emitting surface in the $75\ M_\odot$ model, given the more slowly varying and lower-amplitude spiral SASI accretion and neutrino-emission modulations in this model (because of its large shock radius). In both cases, the dipole never reaches a stationary location. The fact that the $\bar\nu_e$ and ELN emission dipoles are not closely correlated but evolve independently signifies that they do not originate from a single effect, but SASI and LESA both play a role. }
\label{fig:LESAdip}
\end{figure*}

Reference~\cite{Tamborra:2014aua} showed that, once the ELN dipole due to LESA is developed, its direction remains essentially stable (see, e.g., their Fig.~1). Figure~\ref{fig:LESAdip}  shows the evolution of the ELN dipole direction for both BH-forming models. In order to distinguish between LESA and SASI, the $\bar\nu_e$ luminosity dipole direction is also plotted in gray to guide the eye (see also Fig.~\ref{fig:s40_Lum_Dip_Evo} and Fig.~14 of~\cite{Tamborra:2014hga} for comparison). It is evident that the $\bar{\nu}_e$ dipole direction and the ELN dipole direction are not correlated, both develop basically independently. This is a clear signature that both do not originate only from a single (hydrodynamical) phenomenon. While the $\bar{\nu}_e$ emission dipole results mainly from SASI-induced accretion modulations, the ELN dipole is a superposition of effects from SASI and LESA. The non-stationary drifting of the ELN dipole suggests that the SASI-ELN dipole dominates over the LESA-ELN dipole and the migration of the ELN dipole is mainly a consequence of the strong and long-lasting SASI activity. This is especially evident  in the $40\ M_\odot$ model, where the spiral SASI-induced accretion and neutrino-emission modulations are particularly fast (because of the small shock radius) and their relative amplitudes are considerably larger than in the $75\ M_\odot$ case.  

\begin{figure*}
\centering
\includegraphics[width=1.8\columnwidth]{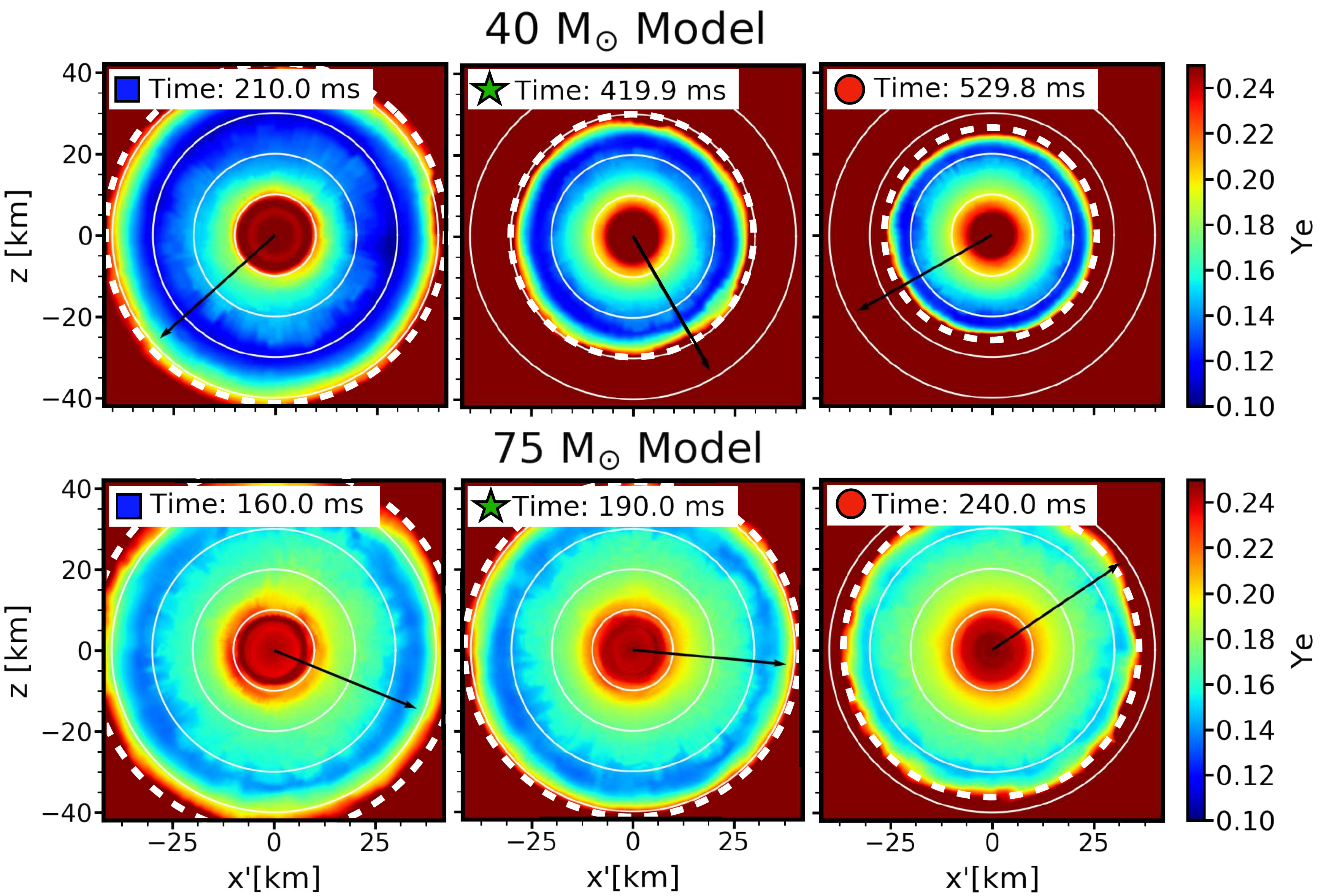}
\caption{Distribution of the electron fraction, $Y_e$, in the PNS for the  $40\ M_\odot$ (top) and $75\ M_\odot$ (bottom) models at the three indicated post-bounce times. For each selected angular direction  (from left to right, see directions marked on Fig.~\ref{fig:LESAdip}), the cut plane is chosen in such a way to contain the ELN dipole vector and the $z$ axis of the polar grid for the data (see Fig.~11 and Sec.~V of \cite{Walk:2019ier} for more details).  The plots are oriented in such a way that the north and the south poles of the data grid are at the top and bottom, respectively. The white circles indicate radii at every $10$~km to guide the eye, and the white dashed circles highlight the location of the PNS radius. The black arrows point in the direction of the positive ELN dipole moment.   A marginally asymmetric shell with slightly stronger deleptonization  in the direction of minimal lepton number flux is present within the PNS convection zone in some snapshots, but the effect is by far not as pronounced as in our previous models.  }
\label{fig:Ye}
\end{figure*}

Finally, the LESA instability is responsible for generating a dipole in the electron fraction ($Y_e$) around the PNS convection layer~\cite{Tamborra:2014aua,OConnor:2018tuw,Vartanyan:2018iah,Glas:2018vcs,Vartanyan:2019ssu,Janka:2016fox}. Figure~\ref{fig:Ye} shows the  evolution of the  electron fraction  distribution in the PNS for the  $40\ M_\odot$ (top) and $75\ M_\odot$ (bottom) models. From left to right, for each of the selected observer directions, the cut plane is chosen in such a way to contain the ELN dipole vector and the $z$ axis of the data grid (see Fig.~11 and Sec.~V of \cite{Walk:2019ier} for more details). The black arrows point in the direction of the positive ELN  dipole moment.  Only a very mildly asymmetric deleptonized shell develops in both models as a  direct signature of the weak LESA activity. Therefore, no clear dipole asymmetry of the $Y_e$ distribution is visible in directional correlation with the ELN dipole vector. Again, this suggests that the non-stationary dipole of the ELN emission is mostly associated with the  SASI accretion emission. 

Clear indicators of LESA asymmetries in the PNS are obviously much weaker than in our previously analyzed models. LESA has been argued to be a consequence of asymmetric PNS convection~\cite{Tamborra:2014aua,Janka:2016fox,Glas:2018vcs}. Hence we interpret our findings as a consequence  of the fact that the PNS convection layer in the BH-forming models is deeper inside the PNS core. This means that the neutrino diffusion time scale from the convective layer through the overlying, massive, dense accretion mantle to the neutrinosphere in the present models is  longer than in lower-mass progenitors with lower accretion rates. This fact may delay or hamper the development of a pronounced LESA emission asymmetry and a correspondingly asymmetric electron distribution in the PNS convection layer. On the other hand, in~\cite{Nagakura:2019tmy,Vartanyan:2019ssu}, it was found that a higher PNS mass is associated with stronger convection, as a consequence of the deeper gravitational potential well generated by the heavier PNS mass. 

The authors of Ref.~\cite{Vartanyan:2019ssu} reported clear signatures of LESA in the neutrino emission of all of their models, with a strong ELN dipole appearing at or shortly after $\sim$200\,ms in all cases. For all  of their models they witnessed a clear correlation between the orientation of the LESA dipole axis and the dipole axis of the $Y_e$ distribution in the convective PNS (in agreement with~\cite{Tamborra:2014aua,Tamborra:2014hga,Walk:2019ier,Walk:2018gaw,OConnor:2018tuw,Janka:2016fox,Glas:2018vcs}), but no strong evidence that  LESA correlates with either the behavior of the shock surface or the accretion rate. We do not consider the findings of~\cite{Vartanyan:2019ssu} to be in conflict with our interpretation. In all of the models of Refs.~\cite{Nagakura:2019tmy,Vartanyan:2019ssu} the mass accretion rate of the PNS has dropped below (in most cases even considerably below) below, 1\,$M_\odot$\,s$^{-1}$ at the time when the ELN dipole amplitude begins to compete with the monopole of the lepton-number emission, either because the star is exploding or because the non-exploding models are 13--15\,$M_\odot$ progenitors with generically lower accretion rates. This is in line with what we had discussed in Refs.~\cite{Tamborra:2014aua,Tamborra:2014hga,Walk:2019ier,Walk:2018gaw} for the lower-mass models presented there.  In contrast, the high-mass models discussed here possess mass accretion rates of at least 1\,$M_\odot$\,s$^{-1}$ until the end of the simulated evolution, and during most of the post-bounce time even significantly higher values (see Fig.~\ref{fig:simulations}, top panel). Therefore the accretion luminosity is considerably larger and the accretion contribution to the anisotropic ELN emission correspondingly bigger in our BH forming models.

\section{Detectable Features in the Neutrino Signal}\label{sec:Detectable}

Neutrinos are amongst the only potentially detectable probes of a massive star collapsing into a BH. In this Section, we  focus our attention on the detectable characteristics of the emitted neutrino signal, and identify which of the physical features discussed above can be directly inferred from measurable quantities. To this purpose, we estimate the event rate detectable  in IceCube, Super-Kamiokande, and in the Deep Underground Neutrino Experiment (DUNE) for the two  BH-forming  models.

\subsection{Neutrino flavor conversions}\label{sec:conversions}

The neutrino signal observed in a detector on Earth is crucially affected by the flavor conversion history along the way.
Flavor conversions are neglected in the neutrino transport equations in SN simulations, under the assumption that the propagation eigenstates coincide with the weak-interaction eigenstates because of the strong matter effect~\cite{Wolfenstein:1977ue}. However, as neutrinos stream away from the SN core and the matter background decreases, flavor conversions become important, see e.g.~\cite{Mirizzi:2015eza,Horiuchi:2017sku} for recent reviews on the topic.

 Adiabatic neutrino flavor conversions occur because of the SN matter background felt by neutrinos, the so-called Mikheev-Smirnov-Wolfenstein (MSW) effect. According to this picture, a detector mainly sensitive to $\bar\nu_e$ will see a neutrino energy distribution roughly given by  $70\%$ of the un-oscillated $\bar\nu_e$ energy distribution and $30\%$ of the un-oscillated  $\bar\nu_x$ distribution for normal mass ordering. On the other hand, for inverted ordering, the detected $\bar\nu_e$ energy distribution will coincide with the un-oscillated $\bar\nu_x$ distribution at the source. Similarly, a detector sensitive to $\nu_e$ will basically detect the un-oscillated $\nu_x$ for normal ordering, and a linear combination of the un-oscillated energy distributions of $\nu_e$ and $\nu_x$ for inverted ordering.

Such a simple picture can be strongly modified in two ways. First, if the radial profile of the matter background has significant stochastic fluctuations, this  would be responsible for affecting the adiabaticity of the flavor conversions, as it happens in the presence of turbulence~\cite{Kneller:2017lqg,Patton:2014lza}. Second, if the neutrino-neutrino refraction is not negligible, which is usually the case at radii smaller then the ones where MSW effects take place. What exactly happens when neutrino-neutrino interactions dominate the flavor evolution history remains poorly understood because of the non-linear nature of the phenomenon.  In particular, it has been recently postulated that neutrino-neutrino interactions occurring in the proximity of the SN core may occur at a ``fast'' rate determined by the neutrino density and possibly lead to flavor equilibration, see e.g.~Refs.~\cite{Chakraborty:2016yeg,Izaguirre:2016gsx,Sawyer:2005jk}. 

Given the current uncertainties on our understanding of flavor conversion of SN neutrinos, we refrain from considering any specific flavor conversion scenario, and instead rely on the un-oscillated neutrino signal. Independently of the exact mixing outcome, the real signal detected on Earth will be an intermediate case between the $\bar\nu_e (\nu_e)$ un-oscillated signal (mimicking what one would detect in the absence of flavor conversions) and the $\bar\nu_x (\nu_x$) un-oscillated signal (mimicking what one would detect under the assumption of full flavor conversions). 

As for the neutrino energy distribution, it was demonstrated in~\cite{Tamborra:2014hga,Keil:2002in} that the un-oscillated neutrino energy distribution of all flavors can be fitted with extremely high  accuracy with a Gamma distribution. Therefore, in the following, we will rely on the neutrino luminosity, mean energy and second energy moment to reconstruct the neutrino fluence expected on Earth and the related event rate.

\subsection{Expected neutrino event rate}\label{sec:Event_Rate}

The neutrino detector currently providing the largest event statistics for a Galactic SN is the IceCube Neutrino Observatory~\cite{Abbasi:2011ss}. This Cherenkov detector works mainly through the inverse-beta-decay ($\bar{\nu}_e + p \rightarrow n \; + \; e^+$)  channel, and is  mainly sensitive to the $\bar{\nu}_e$ flux. We estimate the IceCube event rate for each BH-forming model using the method outlined in Sec.~V of \cite{Tamborra:2014hga} and folding the neutrino energy distribution with the inverse-beta-decay cross section~\cite{Strumia:2003zx}. The event rate is computed assuming an overall background rate  $R_\mathrm{bkgd} = 1.48 \times 10^3~\mathrm{ms}^{-1}$~\cite{Abbasi:2011ss}.

The signal modulations developing in the BH-forming stellar collapse will also be  detectable by  the water Cherenkov detector Super-Kamiokande~\cite{Ikeda:2007sa,Tamborra:2014hga}. Super-Kamiokande has a fiducial volume of $22.5$~kton and is also mainly sensitive to $\bar{\nu}_e$, albeit with less statistics than IceCube. Notably, Super-Kamiokande has the advantage of being virtually background free  for SN neutrino detection ($R_\mathrm{bkgd} = 0$). For the estimation of the expected event rate, we employ the same inverse-beta-decay cross section as for IceCube and assume a $100\%$ detector efficiency for Galactic SNe~\cite{Fukuda:2002uc}.

\begin{figure*}
\centering
\includegraphics[width=1.5\columnwidth]{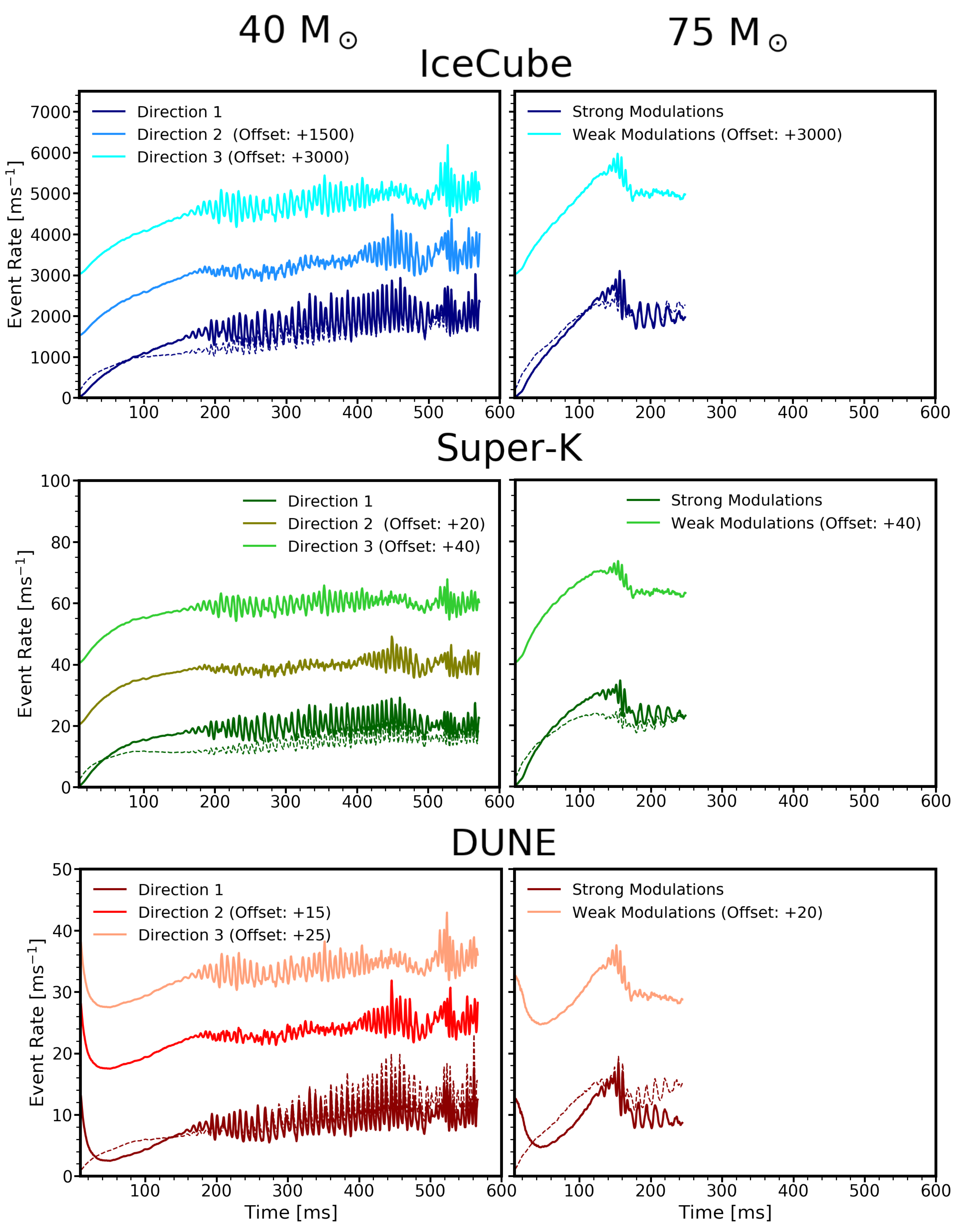}
\caption{Expected detection rate for the $40\ M_\odot$ (left) and $75\ M_\odot$ (right)  models  as a function of post-bounce time for a stellar collapse located at $10$~kpc. The event rate is plotted for IceCube, Super-Kamiokande and DUNE from top to bottom, respectively.  In the top and middle panels, the event rate along Direction 1 has been plotted for $\bar\nu_e$ (continuous line, absence of flavor conversion) and for $\nu_x$ (dashed line, full flavor conversion). The three different colors indicate the observer directions chosen as detailed in Sec.~\ref{sec:Directional_SASI}. For simplicity, the $\bar{\nu}_e$ signal without flavor conversions has been adopted for the other observer directions.  Similarly, the bottom panels show the corresponding event rates in DUNE, for the main $\nu_e-^{40}$Ar detection channel. In all cases, the event rate mirrors the SASI signal modulations for both models.}
\label{fig:IC_Event_Rate}
\end{figure*}

The top panels of Fig.~\ref{fig:IC_Event_Rate} show the predicted IceCube event rate for the $40\ M_\odot$ (left) and $75\ M_\odot$ (right) models for an observer located along the directions selected in Sec.~\ref{sec:Directional_SASI} and for a SN located at a distance of $10$~kpc. Each predicted signal exhibits clear large-amplitude modulations due to spiral SASI for observers located in the proximity of the SASI planes, similar to what was found for the $15$, $20$, and $27\ M_\odot$ models in~\cite{Walk:2018gaw, Tamborra:2014hga, Tamborra:2013laa}. Remarkably, however, due to the strong spiral SASI activity, the modulations in the neutrino signal will also be visible for the $40\ M_\odot$ model for observers located away from the main SASI planes. Additionally, the quadrupolar SASI phase of the $75\ M_\odot$ model in the interval $[140, 175]$~ms  is also clearly detectable along any given observer direction. These features present in the neutrino signal will also be detectable by Super-Kamiokande, as shown in the middle panels of Fig.~\ref{fig:IC_Event_Rate}, although with a reduced event rate. As visible from Fig.~\ref{fig:IC_Event_Rate}, BH-forming stellar collapses are expected to have a slightly increased event rate than ordinary core-collapse SNe  (see e.g.~\cite{Tamborra:2013laa,Tamborra:2014hga,Seadrow:2018ftp,Takiwaki:2017tpe,Kuroda:2017trn} for comparison). 

\begin{figure*}
\centering
\includegraphics[width=2\columnwidth]{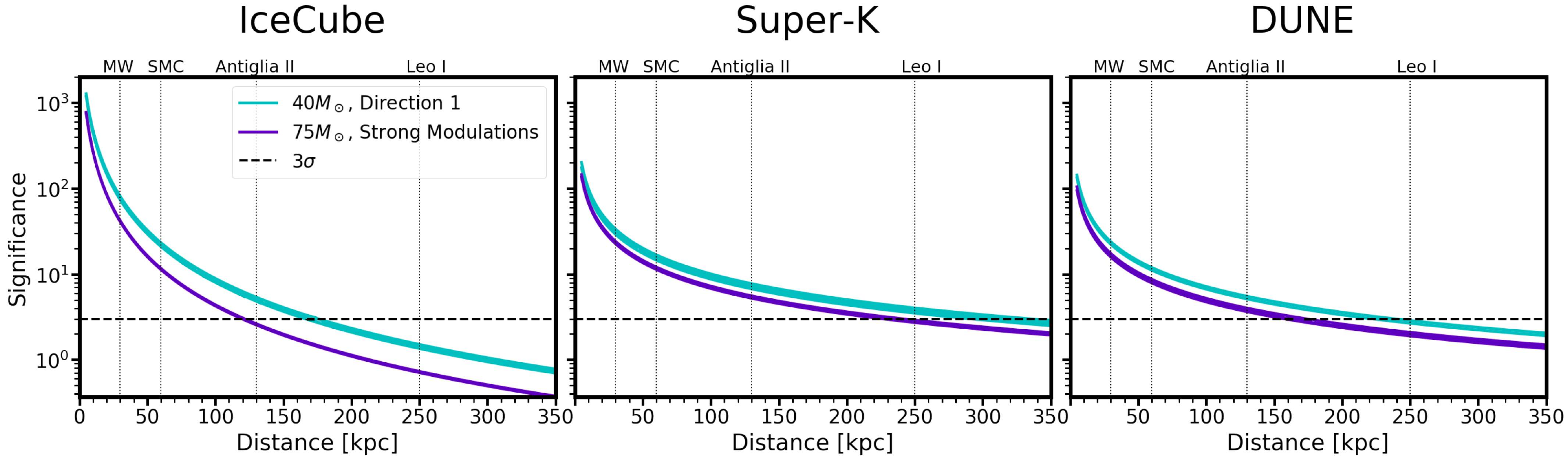}
\caption{Detection significance of the neutrino burst from BH formation for the $40\ M_\odot$ (cyan) and $75\ M_\odot$ (purple) models as a function of the BH-forming collapse event distance for the IceCube Telescope, Super-Kamiokande, and DUNE, from left to right. For both models, the direction yielding the highest number of events has been chosen (i.e., Direction 1 for the $40\ M_\odot$ model, and the Strong Modulations direction for the $75\ M_\odot$ model). The bandwidth in each detection significance curve reflects the two extreme flavor oscillation scenarios. The black dashed horizontal line marks the $3\sigma$ detection significance. The vertical dotted lines mark the edge of the Milky Way (MW) as well as the Small Magellanic Cloud (SMC), Antiglia II, and Leo I. The detection of the neutrino burst with DUNE and Super-Kamiokande will be more likely than for IceCube at large distances since these detectors are virtually background free. } 
\label{fig:detection_sign}
\end{figure*}

 Figure~\ref{fig:detection_sign} shows the detection significance of the time-integrated neutrino burst preceding the BH formation, as a function of distance:  
\begin{eqnarray}
\sigma = \frac{N_{\mathrm{ev}}}{{\sqrt{N_{\mathrm{bkgd}}} + \sqrt{N_{\mathrm{ev}}}}}\ , 
\label{eq:det_sig}
\end{eqnarray}
where $N_{\mathrm{ev}}$ is the total number of SN neutrino events integrated over the signal duration at a chosen  distance from the BH-forming stellar collapse event, and $N_{\mathrm{bkgd}}$ is the total number of background events in the same time window. The left panel of Fig.~\ref{fig:detection_sign} shows that a BH-forming event will be detectable at more than $3\sigma$ by IceCube up to $\mathcal{O}(100)$~kpc. For comparison, the middle panel shows the detection significance of Super-Kamiokande. Here, the burst prior to BH formation will be detectable in neutrinos up to $\mathcal{O}(250)$~kpc due to the absence of  background in Super-Kamiokande over the relatively small time window of a SN burst.
 
Given the strong spiral SASI activity of the models investigated in this paper and the high  event rate for BH-forming models expected from IceCube and Super-Kamiokande, we foresee that the characteristic neutrino signatures will also be visible in the upcoming DUNE neutrino detector~\cite{Acciarri:2015uup}. DUNE is a $40$~kton liquid-argon time projection chamber planned to be in complete operation within 2026. The main detection channel for low energy neutrinos will be the charged current absorption of electron neutrinos on $^{40}$Ar  ($\nu_e + ^{40}$Ar $\rightarrow e^- + ^{40}$K$^\ast$) with an energy threshold of $5$~MeV and  a planned detection efficiency of $86\%$~\cite{Acciarri:2015uup}. 

Similarly to the IceCube and Super-Kamiokande event rates, the DUNE event rate can be estimated by folding the emitted neutrino flux with the $\nu_e$-$^{40}$Ar cross-section provided in~\cite{Acciarri:2015uup} and by taking into account the detection efficiency. The bottom panels of Fig.~\ref{fig:IC_Event_Rate} show the predicted neutrino event rate in DUNE as a function of time  for the $40\ M_\odot$ (left) and $75\ M_\odot$  (right) models along the observer directions selected in Sec.~\ref{sec:Directional_SASI} and at a distance of $10$~kpc. The main background for the detection of  neutrinos from stellar collapse events is thought to be about $122$ solar neutrinos per day~\cite{Acciarri:2015uup}. As it can be seen from the bottom panels of Fig.~\ref{fig:IC_Event_Rate}, this background rate is negligible compared to the event statistics of the neutrino burst from BH-forming stellar collapse events. Thus, we set $R_{\mathrm{bkgd}}=0$, similar to what is considered for Super-Kamiokande. The total expected number of events for the $75\ M_\odot$ model increases by $\simeq 30\%$ on average, if the charged current absorption of $\bar\nu_e$ on $^{40}$Ar and the neutral current scattering of all flavors on $^{40}$Ar are included. The gap between the event rates of $\nu_e$ and the heavy lepton flavors clearly visible in the bottom right panel of Fig.~\ref{fig:IC_Event_Rate}, becomes increasingly large as the post-bounce time increases, because, in this particular model, the emitted flux  of $\nu_e$'s drops continuously after the infall of the Si/O interface while the flux of $\bar\nu_e$ and $\nu_x$ increases. 

As shown in the right panel of Fig.~\ref{fig:detection_sign},  the detection of neutrinos from BH-forming collapse events in DUNE will occur with a significance larger than $3\sigma$ for bursts located up to $170$~kpc for the $75\ M_\odot$ model and $240$~kpc for the $40\ M_\odot$ model.  In fact, although Super-Kamiokande and DUNE will have less statistics than IceCube (see Fig.~\ref{fig:IC_Event_Rate}), the detection of neutrinos from  stellar core collapse events  with these detectors will be more promising than for IceCube at large distances since these detectors are virtually background free.  This feature is very encouraging for what concerns the detection of BH-forming collapse events in neutrinos. 

\subsection{Fourier analysis of the event rate}\label{sec:Fourier}

In this Section, we further identify the detectable features of SASI unique to BH formation by studying the frequency content of the neutrino event rate of the two BH-forming models in Fig.~\ref{fig:IC_Event_Rate}. For this, we investigate the spectrograms of the neutrino event rate obtained as detailed in Sec.~IV of~\cite{Walk:2018gaw}, and the power spectrum of the neutrino event rate, computed following~\cite{Lund:2010kh}. 

\begin{figure*}
\centering
\includegraphics[width=2\columnwidth]{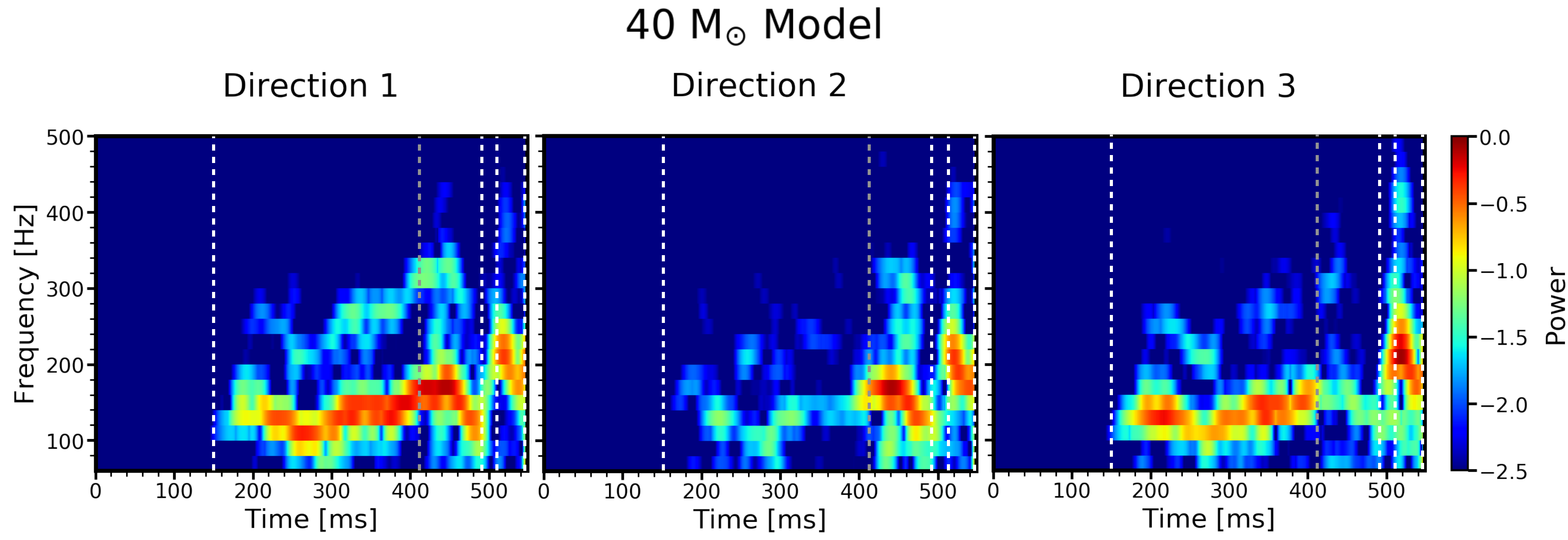}
\caption{Spectrograms of the $\overline{\nu}_e$ IceCube event rates in the absence of flavor conversions for the $40\ M_\odot$ model along the three directions chosen in Sec.~\ref{sec:Directional_SASI}. The spectrogram power has been normalized to the maximum Fourier power along each selected observer direction, and plotted on a log color scale. Dashed vertical lines indicate the same SASI intervals as in Fig.~\ref{fig:s40_Properties}, but extend forward in time by $10$~ms to compensate for the effects of the sliding window over which the short-time Fourier transform is computed. Because of  the 50~ms sliding window, the plotted spectrograms end 25~ms before the end of the simulation. The spiral SASI activity is characterized by hot red regions in the spectrograms. The evolution of the spiral SASI frequency is clearly visible. Negligible activity is found for frequencies larger than 500~Hz.} 
\label{fig:s40_Spec}
\end{figure*}

Figure~\ref{fig:s40_Spec} shows the spectrograms of the IceCube event rate for the $40\ M_\odot$ model along the three observer directions selected in Sec.~\ref{sec:Directional_SASI}. Each spectrogram is normalized to the maximum Fourier power along the selected observer direction. Along Direction 1, the first spiral SASI episode is identifiable through the stripe in hotter colors appearing in correspondence to the spiral SASI frequency [$\mathcal{O}(100)$~Hz]. Remarkably,  given the long-lasting spiral SASI phase, the hot red region of the spectrogram clearly highlights dips and peaks as well as an overall increase of the SASI frequency as a function of time. These trends can be explained by taking into account that the spiral SASI frequency depends on the shock radius and on the NS radius in the following way~\cite{Scheck:2007gw}
\begin{equation}
f_{\mathrm{SASI}}^{-1} \simeq \int_{R_{\mathrm{NS}}}^{R_{\mathrm{s}}} \frac{dr}{|v|} + \int_{R_{\mathrm{NS}}}^{R_{\mathrm{s}}} \frac{dr}{c_s - |v|}\ ,
\end{equation}
where $c_s$ is the radius-dependent sound speed and $v$ is the accretion velocity in the post-shock layer. As can be seen in Fig.~\ref{fig:simulations}, in the $40\ M_\odot$ model, the shock radius contracts reaching a local minimum around $200$~ms, then it slightly expands around $300$~ms until it reaches a stationary value; correspondingly,  $R_{\mathrm{NS}}$ contracts and the post-shock velocity tends to follow a trend opposite to the shock radius (smaller shock radii lead to higher magnitudes of the post-shock velocity and viceversa), see Figs.~\ref{fig:simulations} and \ref{fig:s40_Properties}. Therefore, the SASI frequency tends to decrease during phases of shock expansion and to increase during periods of shock retraction. Thus, in Fig.~\ref{fig:s40_Spec}, one can clearly see that $f_{\mathrm{SASI}}$ tracks the shock contraction and expansion preceding the onset of BH formation,  in agreement with what was diagnosed  in~\cite{Foglizzo:2005xr,Foglizzo:2006fu} and still holds in the case of the spiral SASI. 

Notably, even the spectrogram of the signal along  Direction 2 (i.e., along one of the least optimal directions for observing the modulations in the first sub-interval of the long spiral SASI episode) shows clear signs of the evolution of the spiral SASI frequency in time. The less prominent, but still traceable hot region in the first sub-interval, $[160, 420]$~ms, lines up perfectly with the brighter hot region in the second sub-interval, $[420, 500]$~ms, as it does along Direction 1 and 3. Also, along Direction 3, the colored regions line up at the boundary between the two sub-intervals ($t_{\mathrm{p.b.}} \simeq 420$~ms), indicating that the hot region in $[420, 500]$~ms marks a continuation of the spiral SASI episode along a slightly different direction. 

\begin{figure*}
\centering
\includegraphics[width=1.5\columnwidth]{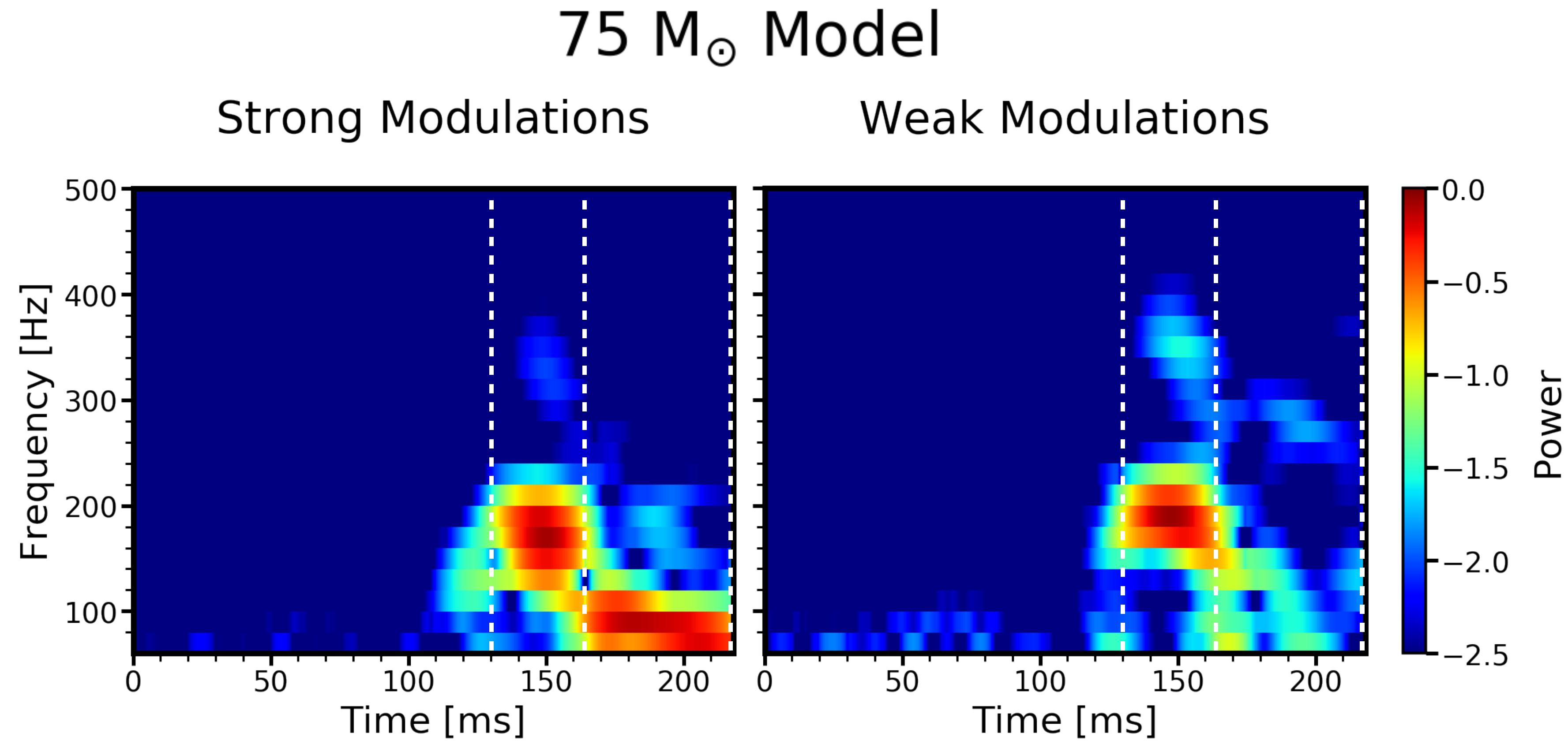}
\caption{Spectrograms of the $\overline{\nu}_e$ IceCube event rates similar to Fig.~\ref{fig:s40_Spec}, but for the $75\ M_\odot$ model. The left panel shows the Strong Modulations direction, which lies along the spiral SASI plane, showing activity corresponding to the dipole spiral SASI frequency in the interval $[175, 230]$~ms. This red  spiral SASI  region nearly disappears along the Weak Modulations direction as expected. The region corresponding to the quadrupolar SASI modulations in the interval $[140, 175]$~ms is directionally independent and present in both spectrograms. The quadrupolar SASI frequency is higher than the dipolar spiral SASI frequency. Dashed vertical lines indicate the same SASI intervals as in Fig.~\ref{fig:u75_Properties}, but extend forward in time by 10 ms to compensate for the effects of the sliding window over which the short-time Fourier transform is computed. Due to the 50~ms sliding window, the plotted spectrograms end 25~ms before the end of the simulation.}
\label{fig:u75_Spec}
\end{figure*}

Figure~\ref{fig:u75_Spec} gives the spectrograms of the IceCube event rate for the $75\ M_\odot$ model along the two observer directions chosen in Sec.~\ref{sec:Directional_SASI}. As expected, the hot red region corresponding to the spiral SASI frequency in the interval $[175, 230]$~ms nearly disappears between the Strong (left) and Weak (right) Modulation directions. The frequency of the quadrupolar SASI modulations  in the IceCube event rate in the interval $[140, 175]$~ms are  represented by a hot red region, visible as expected in both spectrograms due to their directional independence. The left hand panel shows that the spiral SASI dipole frequency is clearly lower than the frequency of the SASI quadrupolar motion. In fact, as visible from Fig.~\ref{fig:simulations},  for the $75\ M_\odot$ model, the relative difference $R_{\mathrm{s}} - R_{\mathrm{NS}}$ grows as a function of time as $R_{\mathrm{s}}$ expands and $R_{\mathrm{NS}}$ contracts, and this is responsible for a drop of the SASI frequency from the quadrupolar to the dipolar phase~\cite{Kazeroni:2015qca}. As previously discussed in Sec.~\ref{sec:Evolution_SASI}, at the transition between the quadrupolar and the dipolar phase, the shock radius shows a contraction  followed by an expansion  (between $130$ and $160$~ms just before the approach of the Si/O interface) that is also tracked by the drop of the SASI frequency in the same time interval. Similar information should also  be contained in the  spectrograms of the correspondent gravitational wave signal~\cite{Srivastava:2019fcb,Takiwaki:2017tpe,Kuroda:2017trn,Andresen:2018aom,Westernacher-Schneider:2019utn}. However, a dedicated analysis of the imprints of BH formation in the gravitational wave signal will be subject of a future  paper.

\begin{figure*}
\centering
\includegraphics[width=2.\columnwidth]{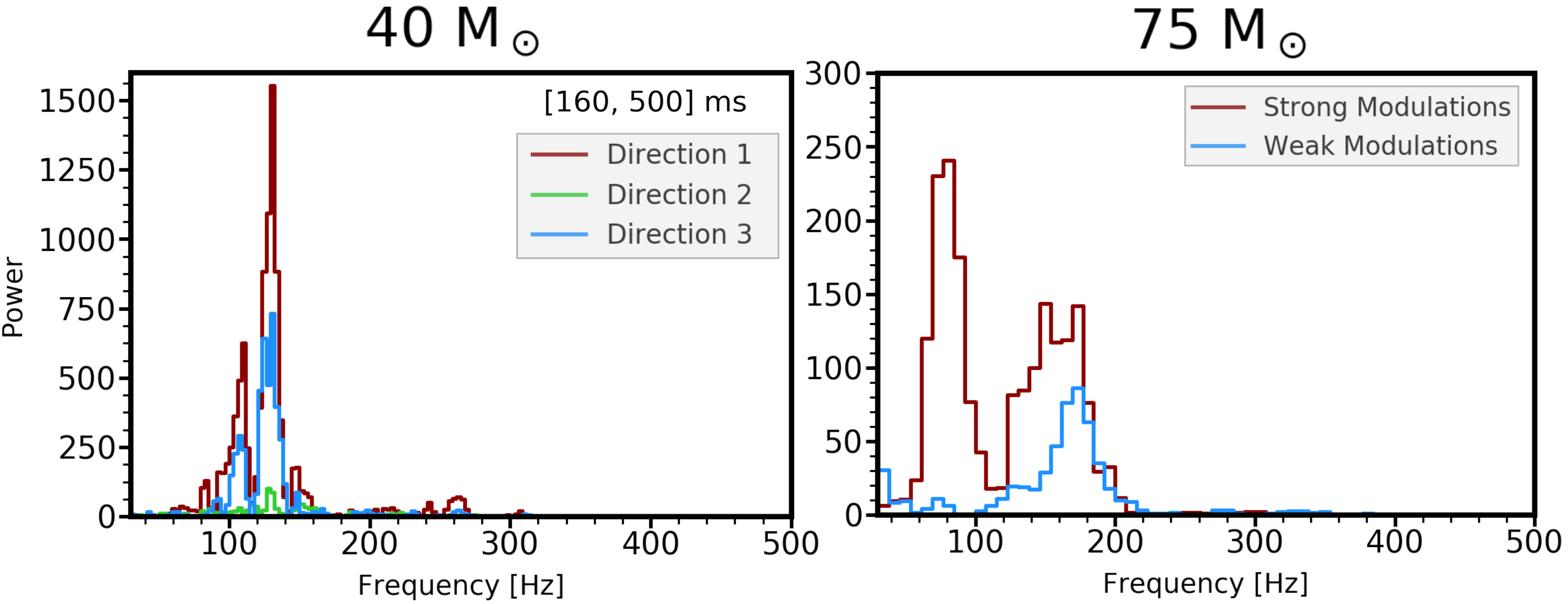}
\caption{Fourier power spectrum of the $\bar{\nu}_e$ IceCube event rate for the $40\ M_\odot$ (left) and the $75\ M_\odot$  models (right) at 10 kpc. Each. power spectrum has been normalized to the average power of the IceCube shot noise. In the left panel, the Fourier spectrum for the $40\ M_\odot$ model is plotted along Direction 1 (dark red), Direction 2 (lime green), and Direction 3 (pale blue) selected as in  Sec.~\ref{sec:Directional_SASI}, over the time interval of the first spiral SASI episode. The two separate peaks indicate the dominant spiral SASI frequency in the modulations of the first and second phases of the spiral SASI episode respectively.  A higher ``overtone'' is visible as low peak around $260$~Hz (see also Fig.~\ref{fig:s40_Spec}). The Fourier power spectrum of the full $\bar{\nu}_e$ IceCube event rate for the $75\ M_\odot$ model is plotted along the directions of strong (dark red) and weak (pale blue) modulations determined in Sec.~\ref{sec:Directional_SASI}. The two peaks correspond to the SASI  quadrupolar and dipolar activity.  The dipolar spiral SASI peak is sharp and prominent, whereas the peak corresponding to the quadrupolar SASI is broader.}
\label{fig:PS_s40}
\end{figure*}
 
Figure~\ref{fig:PS_s40} shows the Fourier power spectra of the IceCube event rate for the $40\ M_\odot$ (left) and the $75\ M_\odot$ model (right), normalized to the power of  a shot noise realization of the IceCube background event rate ($R_\mathrm{bkgd}$). On the left hand side, the power spectrum for the $40\ M_\odot$ model has been computed in the interval of the first spiral SASI episode, $[160,500]$~ms, along each of the three observer directions selected as in  Sec.~\ref{sec:Directional_SASI}. Two different frequency peaks can be clearly identified, one at $\sim 110$~Hz and the other one at $\sim 130$~Hz,   corresponding to the spiral SASI frequency in $[160,420]$~ms and $[420,500]$~ms sub-intervals, respectively.  Thus, there is an increase in frequency of about $20$~Hz as the shock radius retracts. However, this feature will only be detectable along directions where all SASI peaks rise above the power of the shot noise in IceCube. The right panel of Fig.~\ref{fig:PS_s40} shows the Fourier power spectrum of the $75\ M_\odot$ model. Along the Strong Modulation direction, two peaks can be clearly identified; one corresponding to the dipolar spiral SASI frequency at $\sim 80$~Hz, and one corresponding to the frequency of the quadrupolar SASI at $\sim 160$~Hz, higher than the former~\cite{Guilet:2011aa}, as expected due to the contraction of the shock radius. As expected, the peak of the dipole spiral SASI frequency disappears along the Weak Modulations direction. 

\begin{figure*}
\centering
\includegraphics[width=2.\columnwidth]{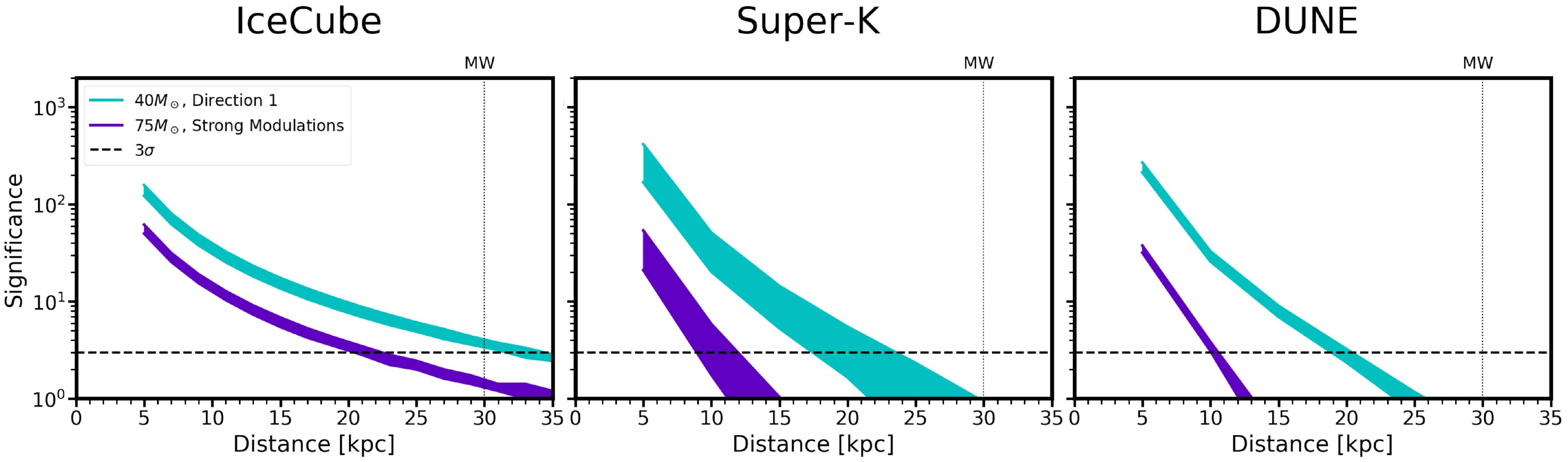}
\caption{Detection significance of the spiral SASI peak in the Fourier spectrum of the IceCube (left), Super-Kamiokande (middle), and DUNE (right) event rates for the $40\ M_\odot$ (cyan) and the $75\ M_\odot$ (purple) models compared to that of the detector noise. Each Fourier decomposition is done for the event rate along the most optimal directions; Direction 1 for the $40\ M_\odot$ model and the Strong Modulations direction for the $75\ M_\odot$  model. The bandwidth in each detection significance curve reflects  the two extreme scenarios absent or complete flavor conversion. The black dashed horizontal line marks the $3\sigma$ detection significance. The vertical dotted lines mark the edge of the Milky Way. The SASI frequency will be detectable above the detector noise within our own Galaxy.}
\label{fig:PS_sigma}
\end{figure*}

Figure~\ref{fig:PS_sigma} presents the detection significance of the spiral SASI peak normalized to the power of the detector noise in the Fourier spectrum of the IceCube, Super-Kamiokande, and DUNE event rates for the  $40\ M_\odot$ and the $75\ M_\odot$  models as a function of the SN distance.  As discussed in the previous Section, the IceCube noise model includes the shot noise of the detector and statistical fluctuations of the incoming neutrino signal. In Super-Kamiokande and DUNE, which are considered background free over the short time interval of the stellar collapse, the noise model includes only statistical Poissonian fluctuations of the detected neutrino signal. For each model, we choose the direction yielding the highest number of total events, i.e, Direction 1 for the $40\ M_\odot$  model, and the Strong Modulations direction for the $75\ M_\odot$ model. The band is found by considering the two  scenarios of absence of flavor conversions or  full flavor conversion, as shown in Fig~\ref{fig:IC_Event_Rate}. In IceCube, the spiral SASI peak will be distinguishable over the detector noise for distances up to 35~kpc for the $40\ M_\odot$ model, and $\sim$~22~kpc for the $75\ M_\odot$ model, while it will  be detectable in DUNE and Super-Kamiokande only up to distances of $\mathcal{O}(10-20)$~kpc.
 
In conclusion, due to the higher neutrino event statistics, the Fourier power spectrum of the IceCube event rate will be able to determine the SASI frequency at greater distances than DUNE and Super-Kamiokande. However, due to the absence of background in DUNE and Super-Kamiokande, these detectors will be able to detect the neutrino burst from  a BH-forming stellar collapse event  up to impressively large distances.

\section{Conclusions}\label{sec:Conclusions}

Intriguingly, very little is known about the properties of black hole (BH) forming stellar collapse events. Throughout this work, we aim to provide a first attempt to infer detectable characteristics of the neutrino signal unique to BH formation, by exploring the neutrino emission properties of the 3D hydrodynamical simulations of two BH-forming progenitors with different masses ($40\ M_\odot$ and  $75\ M_\odot$) and metallicities.  

The two models have different BH formation timescales ($\simeq 570$~ms for the $40\ M_\odot$ model and $250$~ms for the $75\ M_\odot$ model). Interestingly, while the $75\ M_\odot$ model exhibits a shock expansion before the collapse into a BH, similarly to what was found in~\cite{Chan:2017tdg,Kuroda:2018gqq,Pan:2017tpk}, the shock radius evolves towards a quasi-stationary value in the $40\ M_\odot$ model.  The extremely high accretion rate causes the neutrino  luminosity to be dominated by accretion luminosity, and   the ELN emission dipole and quadrupole are dominated by the rapidly fluctuating spiral SASI-induced modulations of amplitude and direction.  The variations of the time-averaged (and in this sense stable) ELN emission with direction, relative to the total neutrino number flux, are considerably weaker  than in previous ``normal'' core-collapse models, and the anti-correlation of the $\nu_e$ and $\bar\nu_e$ emission variations is much smaller or barely visible. The same holds true for asymmetries of the electron distribution around the proto-neutron star convection layer. All together,  the implications of the  neutrino-driven self-sustained asymmetry (LESA) are minor in the BH-forming stellar collapses investigated in this work.

Extremely strong and long-lasting  spiral and quadrupolar SASI episodes occur in these models. Since the neutrino event rate for BH-forming stellar  collapse is expected to be higher than that of ordinary core-collapse supernovae because of the higher luminosity emitted in neutrinos, the spiral SASI frequency will be detectable by the IceCube Neutrino Telescope for BH formation  occurring up to  distances of 35~kpc ($\sim$~22~kpc) for the $40\ M_\odot$ ($75\ M_\odot$) models. Similarly, the detection prospects are limited to our own Galaxy for  Super-Kamiokande and DUNE. 

Notably, given the long-lasting spiral SASI, the evolution of the spiral SASI frequency  will be clearly visible, e.g.~in the spectrogram of the IceCube event rate. Moreover, SASI imprints will be detectable in neutrinos even for observes located away from the spiral SASI plane because of the  strong SASI activity. 

The two BH-forming progenitor models explored throughout this work illustrate the phenomenal power of using neutrinos to study the physical processes involved in BH formation. The excellent detectability prospects in neutrinos for these yet mysterious astrophysical events have the potential to unveil the inner workings of collapsing massive stars.

\section*{Acknowledgments}

We are grateful to Bernhard M{\"u}ller for useful comments on the manuscript and Tobias Melson for providing access to the data set adopted in this paper. This project was supported by the Villum Foundation (Project No.~13164), the Danmarks Frie Forskningsfonds (Project No.~8049-00038B), the Knud H\o jgaard Foundation, the European Research Council through grant ERC-AdG No.\
341157-COCO2CASA, the Deutsche Forschungsgemeinschaft through Sonderforschungbereich
SFB~1258 ``Neutrinos and Dark Matter in Astro- and
Particle Physics'' (NDM), and under Germany's Excellence Strategy through the Excellence Cluster ``ORIGINS: From the Origin of the
Universe to the First Building Blocks of Life''
(EXC~2094--39078331). The model calculations were performed on SuperMUC at
the Leibniz Supercomputing Centre with resources granted
by the Gauss Centre for Supercomputing (LRZ project ID: pr74de).

\bibliography{BH_Progenitors}

\end{document}